\documentclass[prb,twocolumn,showpacs]{revtex4-1}
\usepackage[T1]{fontenc}
\usepackage{graphicx}
\usepackage{color}
\usepackage[dvipsnames]{xcolor}
\usepackage{dcolumn}
\usepackage{amsmath}
\usepackage{amssymb}
\usepackage{amsfonts,dsfont,xfrac}
\usepackage[bbgreekl]{mathbbol}
\usepackage{subfigure}
\usepackage{braket}
\usepackage{simplewick}

\usepackage{mathrsfs}
\usepackage{relsize,scalerel}
\usepackage{makecell}
\usepackage[colorlinks=true, citecolor=black, linkcolor=black, urlcolor=black]{hyperref}
\usepackage{cleveref}

\newcommand{\eq}{\sss{\textup{eq}}}

\newcommand{\Bcal}{\mathcal{B}}

\newcommand{\Ical}{\mathcal{I}}

\newcommand{\Ncal}{\mathcal{N}}

\newcommand{\Ocal}{\mathcal{O}}

\newcommand{\bR}{\boldsymbol{R}}

\newcommand{\bD}{\boldsymbol{D}}

\newcommand{\bu}{\boldsymbol{u}}
\newcommand{\br}{\boldsymbol{r}}

\newcommand{\bq}{\boldsymbol{q}}
\newcommand{\bk}{\boldsymbol{k}}

\newcommand{\bPsi}{\boldsymbol{\Psi}}

\newcommand{\sss}[1]{\scriptscriptstyle{{#1}}}

\newcommand{\VKS}{V^{\scriptscriptstyle{\text{KS}}}}

\newcommand{\Hel}{\hat{H}{}_{\scriptscriptstyle{\text{el}}}}

\newcommand{\Nc}{N_{{c}}}

\newcommand{\gn}[3]{\overset{\sss{(#1)}}{g}{}_{#2}^{#3}}

\newcommand{\Avg}[1]{\left\langle{#1}\right\rangle}
\newcommand{\AAvg}[1]{\left\langle\!\!\!\left\langle{#1}\right\rangle\!\!\!\right\rangle}
\newcommand{\ApAvg}[1]{\left\langle\!\!\left\langle{#1}\right\rangle\!\!\right\rangle}

\newcommand{\avg}[1]{\langle{#1}\rangle}
\newcommand{\bReq}{\bR_{\eq}}

\newcommand{\osss}[2]{\overset{\scriptscriptstyle{#1}}{#2}}
\newcommand{\ssst}[1]{\scriptscriptstyle{\text{#1}}}

\newcommand{\elph}{\ssst{elph}}
\newcommand{\el}{\ssst{(el)}}
\newcommand{\nuc}{\ssst{(\negthinspace nu \hspace {-3pt})}}

\newcommand{\Velph}{W^{\ssst{ph}}}
\newcommand{\Velphhat}{\hat{W}{}^{\ssst{ph}}}

\newcommand{\Vn}{V^{\ssst{nu}}}
\newcommand{\Vnnot}{\overset{\sss{(0)}}{V}{}^{\ssst{nu}}}    
\newcommand{\Venot}{\overset{\sss{(0)}}{V}{}^{\ssst{el}}}    
\newcommand{\Ve}{V^{\ssst{el}}}
\newcommand{\Eel}{E^{\ssst{el}}}
\newcommand{\Vks}{V^{\ssst{KS}}}
\newcommand{\Vkshat}{\hat{V}{}^{\ssst{KS}}}
\newcommand{\whVks}{\hat{V}{}^{\ssst{KS}}}
\newcommand{\hVks}{\hat{V}{}^{\ssst{KS}}}
\newcommand{\neeq}{n_{\ssst{eq}}}

\newcommand{\asqF}{{\alpha\!}^2\!F}
\newcommand{\ef}{\epsilon_{\sss{F}}}

\renewcommand{\Im}{\text{Im}\,}
\renewcommand{\Re}{\text{Re}\,}
\newcommand{\FS}{\text{FS}}

\newcommand{\Avgeq}[1]{\Avg{#1}_{\eq}}
\newcommand{\avgeq}[1]{\avg{#1}_{\eq}}
\newcommand{\rhoeq}{\rho_{\eq}}

\newcommand{\Conv}{%
  \mathop{\scalebox{1.5}{\raisebox{-0.2ex}{$\circledast$}}
  }
}

\newcommand{\gavg}{\bigl\langle\mkern-1mu g \mkern-1mu \bigr\rangle{}}
\newcommand{\gavgx}[1]{\bigl\langle\mkern-1mu \overset{\sss{(#1)}}{g} \mkern-1mu \bigr\rangle{}}
\newcommand{\gavgn}{\gavgx{n}}

\newcommand{\gx}[1]{\overset{\sss{(#1)}}{g}{}}
\renewcommand{\gn}{\gx{n}}

\newcommand{\pedReq}{{\strut_{\scriptscriptstyle{\bReq}}}}

\newcommand{\alphasqF}{\alpha^2\!F}

\newcommand{\Wen}{W^{\ssst{ph}}}
\newcommand{\Wenn}{\overset{\sss{(n)}}{W}{}^{\ssst{ph}}}

\newcommand{\GWen}{GW^{\ssst{ph}}}

\newcommand{\lambdax}[1]{\overset{\sss{(#1)}}{\lambda}}

\newcommand{\buI}{{\bu}_{\sss{(\Ical)}}}

\newcommand{\NI}{\Ncal_{\sss{(\Ical)}}}
\newcommand{\bRI}{\overset{\sss{(\Ical)}}{\bR}}
\newcommand{\omegaD}{\omega_{\ssst{D}}}
\newcommand{\Nf}{N_{\sss{F}}}

\newcommand{\enex}[1]{\varepsilon_{#1}}
\newcommand{\ene}{\varepsilon}

\newcommand{\kB}{k_{\ssst{B}}}

\newcommand{\hA}{\hat{A}}
\newcommand{\hB}{\hat{B}}

\newcommand{\hu}{\hat{u}}
\newcommand{\hrho}{\hat{\rho}}
\newcommand{\whrho}{\hat{\rho}}
\newcommand{\hbu}{\hat{\bu}}

\newcommand{\hbR}{\hat{\bR}}
\newcommand{\whbR}{\hat{\bR}}
\newcommand{\wh}[1]{\hat{#1}}
\newcommand{\h}[1]{\hat{#1}}

\newcommand{\hW}{\h{W}}
\newcommand{\whpsi}{\wh{\psi}}
\newcommand{\Tr}[1]{\textup{Tr}\left[#1\right]}

\newcommand{\hWen}{{\hW}{}^{\ssst{ph}}}
\newcommand{\Hnu}{\wh{H}{}^{\mkern-1mu\nuc}}
\newcommand{\Tnu}{\wh{T}{}^{\mkern-1mu\nuc}}
\newcommand{\Vneq}{V_{\ssst{eq}}^{\mkern-1mu\nuc}}
\newcommand{\Vnu}{V^{\mkern-1mu\nuc}}

\newcommand{\bhxi}[1]{\boldsymbol{\hat{\bi}}}

\begin{document}
%\title{Non-perturbative theory of the electron-phonon coupling and its first-principles implementation: the (harmonic/normal/quadratic/Gaussian?)~$G\Wen$~approximation}
\title{Non-perturbative theory of the electron-phonon coupling and its first-principles implementation}

\author{Raffaello Bianco$^{1,2,3,4}$}
\email{raffaello.bianco@unimore.it}
\author{Ion Errea$^{1,5,6}$}
\email{ion.errea@ehu.eus}

\affiliation{$^1$Centro de F\'isica de Materiales (CFM-MPC), CSIC-UPV/EHU,  Manuel de Lardizabal pasealekua 5, 20018 Donostia/San Sebasti\'an, Spain}

\affiliation{$^2$ Ru\dj er Bo\v{s}kovi\'c Institute, 10000 Zagreb, Croatia}

\affiliation{$^3$ Dipartimento di Scienze Fisiche, Informatiche e Matematiche, Universit\`a di Modena e Reggio Emilia, Via Campi 213/a I-41125 Modena, Italy}

\affiliation{$^4$ Centro S3, Istituto Nanoscienze-CNR, Via Campi 213/a, I-41125 Modena, Italy}

\affiliation{$^5$Fisika Aplikatua Saila, Gipuzkoako Ingeniaritza Eskola, University of the Basque Country (UPV/EHU), Europa Plaza 1, 20018 Donostia/San Sebasti\'an, Spain}

\affiliation{$^6$Donostia International Physics Center (DIPC), Manuel Lardizabal pasealekua 4, 20018 Donostia/San Sebasti\'an, Spain}

\begin{abstract}
The harmonic approximation of ionic fluctuations and the linear coupling between phonons and electrons provide the basic approach to compute, from first principles, the contribution that the nuclei dynamics and its interaction with electrons have on materials' properties.  
These approaches are questionable, at least, whenever quantum and anharmonic effects on the nuclei vibrational properties are large, such as in hydrogenous systems, high-$T_c$ superconductors, and when systems are close to charge-density wave or ferroelectric displacive phase transitions. Here we propose a novel non-perturbative approach to compute the electron-phonon interaction from first principles that includes non-linear effects and takes into account the quantum nature of nuclei. The method is based on the $G\Wen$ approximation for the electron self-energy, 
given by the effective nuclei-mediated electron-electron interaction $\Wen$ and the electron Green’s function $G$. 
The electrons are treated at a mean-field level and the nuclei dynamics is described with a Gaussian distribution function, which can effectively take into account anharmonic effects at a mean-field level, e.g. within the self-consistent harmonic approximation. The pivotal quantities of the Gaussian $G\Wen$ self-energy are the Debye-Waller-renormalized average vertices, which are computed in supercells with a stochastic approach, using the the self-consistent electronic potential computed for different atomic configurations. In order to validate the method, $G\Wen$ calculations are performed on aluminum, a highly harmonic system with weak electron-phonon coupling. As expected, the obtained results coincide with the ones obtained with standard linear electron-phonon calculations. However, calculations performed on palladium hydride, a very anharmonic system, show a highly non-linear electron-phonon interaction, with $G\Wen$ bringing  corrections as high as the linear-order result. The performed analysis shows that the developed method may have a large impact on the \textit{ab initio} calculations of all the properties related to the electron-phonon interaction, e.g. superconductivity and electrical conductivity, in highly anharmonic systems.
\end{abstract}

%\pacs{74.25.Kc,74.90.+n,67.63.-r,67.80.F-,63.20.Ry,31.15.A-}

\maketitle

\section{Introduction}

The interaction between electrons and nuclei is, together with the interaction of electrons and nuclei between themselves, one of the cornerstones in condensed matter physics. Indeed, the electron-nuclei interaction (a.k.a. the electron-phonon interaction) determines several physical phenomena like the electrical conductivity in metals, the mobility in doped semiconductors, the temperature dependence of the electronic bands and of the band gap in semiconductors, the optical absorption in indirect-gap semiconductors, and the distortion of phonon dispersions in metals through kinks and Kohn anomalies. Moreover, the electron-nuclei interaction is the key mechanism at the basis of conventional superconductivity, one of the most fascinating emergent phenomena in condensed matter~\cite{RevModPhys.89.015003}. This motivates the unrelenting efforts made in the last decades to implement more and more accurate and efficient approaches to include the effects of the electron-nuclei interaction in the calculation of the electronic properties from first principles. Following Allen-Heine-Cardona (AHC), in the standard approach the starting point is the Born-Oppenheimer approximation, with the two sub-systems (electrons and nuclei) uncoupled, the electron dynamics studied with clamped nuclei and the nuclei dynamics studied at the harmonic level. On top of that, the interaction between electrons and phonons is introduced as a perturbation, given by the variation of the electronic potential when the nuclei are displaced from the equilibrium position, which is treated at the lowest order in the parameter $u/L$ (nuclei displacement $u$ over electronic length-scale $L$)~\cite{PBAllen_1976,PhysRevB.23.1495,PhysRevB.27.4760}. 

Assuming the perturbation at the lowest order in $u/L$ has a twofold implication: on the one hand, it means that the perturbation is defined by  retaining only the lowest-order term (the linear one, typically) in the expansion of the electronic potential w.r.t the atomic displacement; on the other hand, it means that only the first-order correction caused by this perturbation is considered.  In most cases, both these approximations are sensible. However, if very light atoms are present, if temperature is high, or if harmonic restoring forces are very small (e.g. near a second-order phase transition) higher-order terms in atomic displacement (i.e. multiphonon processes) may be significant~\cite{allen1983theory}. In particular, in strongly anharmonic systems the conventional linear approach to the electron-phonon interaction (EPI) is expected to be inadequate. Moreover, there are systems where a strong electron-phonon coupling is measured. For example, high-temperature superconductors are systems where both strong electron-phonon coupling and anharmonicity often coexist~\cite{ALLEN19761157,Lanzara2001,N.M.Plakida_1987,PhysRevB.36.7115}. Indeed, the impact of anharmonicity/multi-phonon processes on superconductivity has been often analyzed in the literature, not only by including higher-order terms from perturbation theory in the calculation of the EPI~\cite{PhysRevB.45.5052}, but also proposing generalizations of the standard electron-phonon formula to compute it, where matrix elements between anharmonic excited states of the nuclei are considered at linear level w.r.t. the displacement and beyond~\cite{Hui_1974,PhysRevLett.60.2191,PhysRevB.48.398,PhysRevLett.87.037001,PhysRevLett.48.264,PhysRevLett.62.2869,PhysRevLett.65.1675,PhysRevB.54.9372,PhysRevLett.68.3236,PhysRevB.56.5297,PhysRevLett.67.228,PhysRevB.56.8322,PhysRevB.47.8050,Chen2022Stochastic,Liu2020Superconducting}. However, a general method to compute the EPI at a non-linear level, not based on models or simplified assumptions, and suited to be used in calculations from first principles that provides a control on the interactions included, is still lacking. 
Moreover, superconductors are not the only materials where the electron-phonon interaction and anharmocity play a dominant role in the physics. Indeed, other systems such as perovskites~\cite{PhysRevLett.64.2575,doi:10.1021/acs.jpclett.6b02560,https://doi.org/10.1002/adma.201704737,doi:10.1021/acs.chemmater.7b01184},  layered and low-dimensional systems~\cite{PhysRevB.87.020506,PhysRevLett.116.127402,GRIONI2004417,PhysRevLett.98.036801,PhysRevB.98.241412}, and  polar materials~\cite{Lee2020} also display a large interplay between these effects. Therefore, the development of an approach to compute EPI effects from first principles at a non-linear level is expected to have a broad range of applications. 

The aim of this work is to present a novel approach to study the EPI at a non-linear level, suited to be used in first-principles calculations. The method, non-perturbative  and based on the Born-Oppenheimer approximation, does not rely on a plain stochastic evaluation of the electron Green function~\cite{PhysRevB.105.245120}, but takes as starting point the expression of the effective nuclei-mediated electron-electron interaction $\Wen$, in connection with the theoretical work presented in Ref.~\citenum{osti_6556355}. The electron self-energy due to this interaction is evaluated at the $G\Wen$ level, the first term in the expansion of the self-energy operator in terms of the effective electron-electron interaction $\Wen$ and the electron Green's function $G$. Therefore, the approach is similar in spirit to the $GW$ approximation for the electron self-energy due to the screened Coulomb electron-electron interaction $W$. Electrons are considered at the Kohn-Sham mean-field level, and the nuclei dynamics is described by an effective harmonic Hamiltonian (i.e. the nuclei have a Gaussian dynamics), which allows to include anaharmonic effects at mean-field level through the usage of, for example,  the self-consistent harmonic approximation (SCHA)~\cite{PhysRevB.89.064302,PhysRevB.96.014111,PhysRevB.98.024106}. Indeed, the method that we are going to present is especially suited to be used in conjunction with the stochastic self-sonsistent harmonic approximation (SSCHA) method.~\cite{Monacelli_2021} In order to evaluate the harmonic $G\Wen$ self-energy, we present an approach based on the stochastic evaluation of the Debye-Waller-renormalized averaged electron-phonon vertices, which requires only the evaluation in supercells of the self-consistent electronic potential for different atomic configurations, and of the electronic wave functions for the undistorted atomic structure.

The paper is structured as follows. In section~\ref{sec:The_nuclei-mediated_electron-electron_interaction}, we derive the analytic expression for the nuclei-mediated effective Kohn-Sham electron-electron interaction $\Wen$. We also collect in table~\ref{tab:symbols}, as a reference for the reader, the main quantities used in the paper, with the corresponding symbol, definition and description. In section~\ref{sec:the_linear_order}, we show that, at the lowest order in the atomic displacement, the interaction $\Wen$ returns the standard single-phonon mediated electron-electron interaction. In section~\ref{sec:nuclei_with_harmonic_dynamics}, we derive the expression of $\Wen$ when the nuclei have Gaussian dynamics. In section~\ref{sec:nuclei_with_harmonic_dynamics_PDA}, we perform a diagrammatic analysis of the interaction $\Wen$ in the Gaussian case, so as to establish the connection of the developed method with the standard perturbative approach to the EPI. In section~\ref{sec:GWen_approx}, we define the (Gaussian) $G\Wen$ approximation. In section~\ref{sec:the_first-principles_implementation}, we describe the first-principles implementation of the described theory. In the subsequent two sections, in order to validate the presented method and demonstrate its capabilities, we show the results of calculations performed from first-principles. In section~\ref{sec:a_case_of_linear_electron-phonon_coupling:_aluminum}, we compute the electron-phonon coupling for aluminum, a system with weak electron-phonon coupling, and where non-linear EPI effects are negligible. In section~\ref{sec:a_case_of_strongly_non-linear_electron-phonon_coupling:_palladium_hydride}, EPC calculations are performed for PdH, a strongly anharmonic system. In section~\ref{sec:conclusions}, we summarize our results and draw some final conclusions. The
paper is completed with several appendices including the proofs of some of the equations given in the manuscript.

% 
% Moreover, there are many cases where strong electron-phonon interaction has been observed, like in perovskites~\cite{PhysRevLett.64.2575,aaa,https://doi.org/10.1002/adma.201704737,doi:10.1021/acs.chemmater.7b01184}, layered and low-dimensional systems~\cite{PhysRevB.87.020506,PhysRevLett.116.127402,GRIONI2004417,PhysRevLett.98.036801,PhysRevB.98.241412}, and polar materials~\cite{Lee2020}. In particular, strong electron-phonon coupling is ubiquitous in high-temperature superconductors~\cite{Lanzara2001}, and  strong anharmoncity is often observed (in particular, in hydrogen based superconductors)
% XXXXXXXXXX
% Several approaches have been adopted to go beyond the standard aproximation used for the EPI, like including higher order terms from perturbation theory~\cite{Lee2020,PhysRevB.45.5052}, or the anharmonic generalization of the electron-phonon coupling obtained including matrix elements over all of the phonon excited states~\cite{Hui_1974,PhysRevLett.60.2191,PhysRevB.48.398,PhysRevLett.87.037001}, or using a stochastic approaches to compute the effect of the atomic fluctuations on electron selfenergy thourgh averages~\cite{PhysRevB.105.245120}.

%%%%%%%%%%%%%%%%%%%%%%%%%%%%%%%%%%%%%%%%%%%%%%%%%%%%%%%%%%%%%
\section{The nuclei-mediated electron-electron interaction}
\label{sec:The_nuclei-mediated_electron-electron_interaction}
In order to include non-linear effects in the electron-phonon coupling, we are going to explictly consider the time-dependent nuclei-mediated electron-electron interaction.
The starting point of our approach is the Born-Oppenheimer approximation with the electrons studied within the Kohn-Sham theory: electrons and nuclei dynamics are decoupled, and when the nuclei are frozen in a configuration $\bR=(\bR_1,\ldots,\bR_N)$, $N$ being the number of atoms,  at equilibrium a ``Kohn-Sham electron'' in $\br$ feels the Kohn-Sham potential $\Vks(\br,\bR)$. More in general, the nuclei are not simply locked in a single point-like configuration, but are described by a time-dependent positional probability distribution $\rho(\bR,t)$. If we keep considering the electronic screening istantaneous (i.e. much faster than the nuclei screening), it seems natural to generalize the potential felt by the electron $\Ve(\br,t)$ through a functional dependence from $\rho(\bR,t)$:
\begin{equation}
%\Ve(\br,t)=\int\,\Vks(\br,\bR)\,\rho(\bR',t)\,d\bR'\,,
\Ve(\br,t)=\int\,\Vks(\br,\bR')\,\rho(\bR',t)\,d\bR'\,,
\label{eq:Ve}
\end{equation}
which of course returns the simple Kohn-Sham potential $\Vks(\br,\bR)$ if $\rho(\bR',t)=\delta(\bR-\bR')$. 
Eq.~\eqref{eq:Ve} is our general definition of the potential felt by the electrons.

Now we consider the nuclei-mediated effective electron-electron interaction. As electrons and nuclei are coupled through the Coulomb interaction, an electron in $\br'$ at time $t'$ causes a lattice distortion that propagates in space and time, which in turn interacts with an electron in $\br$ at time $t$. The nuclei-mediated effective electron-electron potential $\Velph(\br,t,\br',t')$ is nothing but the variation of the electronic potential felt by the electron in $(\br,t)$ due to the fact that another electron was in $(\br',t')$. Therefore, in a linear-response approach, we have that starting from an equilibrium configuration, with time-independent electron and nuclei density $\neeq(\br)$ and $\rhoeq(\bR)$, respectively, a variation of the electronic density $\delta n(\br,t)$  gives a corresponding variation of the electronic potential  $\delta \Ve(\br,t)$ of the form
\begin{equation}
\delta \Ve(\br,t)=\iint\,d\br'\,d t'\,\Velph(\br,t,\br',t')\,\delta n(\br',t')\,.
\label{eq:deltaVe}
\end{equation}
Indeed, notice that considered a test charge in  $(\br',t')$, i.e. $\delta n(\br,t)=\delta(\br-\br')\delta(t-t')$, 
from Eq.~\eqref{eq:deltaVe} we correctly have $\delta \Ve(\br,t)=\Velph(\br,t,\br',t')$. Therefore, we can write
\begin{align}
\Velph(\br,t,\br',t')&=\frac{\delta \Ve(\br,t)}{\delta n(\br',t')}\\
&\mkern-40mu=\iiiint d\bR\,d\bR'\,d\tau\,d\tau'\,\nonumber\\
    &\,\frac{\delta \Ve(\br,t)}{\delta \rho(\bR,\tau)}
    \frac{\delta \rho(\bR,\tau)}{\delta \Vn(\bR',\tau')}
    \frac{\delta \Vn(\bR',\tau')}{\delta n(\br',t')}    ,
  \label{eq:Velph1}
\end{align}
where we have introduced $\Vn(\bR,t)$, the potential felt by the nuclei in $\bR$ at time $t$ (at equilibrium this potential is time independent and is given by the electronic ground-state energy $\Eel(\bR)$  for the considered clamped-nuclei configuration, i.e. $\Vneq(\bR,t)=\Vneq(\bR)=\Eel(\bR)$). 

\renewcommand{\arraystretch}{0.1}
\begin{table*}
\begin{ruledtabular}  
\begin{tabular}{l@{\hskip 0.4cm}l}
   Symbol  & Meaning   \\
\hline
$\phantom{a}$\vspace{0.0cm}\\
$\displaystyle\vphantom{\int}\bR$  
    & Nuclei configuration
    \\  
$\displaystyle\vphantom{\int}\whbR$  
    & Nuclei configuration operator 
    \\  
$\displaystyle\vphantom{\int}\whrho(\bR)=\delta(\bR-\whbR)$  
    & Nuclei density operator  
    \\   
$\displaystyle\vphantom{\int}\Hnu=\Tnu+\Vneq(\bR)$  
    & Unperturbed (i.e. at equilibrium) nuclei Hamiltonian
    \\   
$\displaystyle\vphantom{\int}\whbR(t)=e^{i\Hnu \hspace {-1pt} t}\whbR\,\, e^{-i\Hnu \hspace {-1pt} t}$  
    & Heisenberg evolution of the nuclei configuration operator 
    \\ 
$\displaystyle\vphantom{\int}\whrho(\bR,t)=e^{-i\Hnu \hspace {-1pt} t}\whrho(\bR)\,e^{-i\Hnu \hspace {-1pt} t}
               =\delta(\bR-\whbR(t))$  
    & Heisenberg evolution of the nuclei density operator 
    \\ 
$\displaystyle \h{\Gamma}{}_{\eq}=e^{-\beta\Hnu}/\Tr{e^{-\Hnu}}$
    & Unperturbed nuclei density matrix operator
    \\    
$\displaystyle\vphantom{\int}\avg{\wh{\Ocal}}_{\eq}=\Tr{\h{\Gamma}{}_{\eq}\,\wh{\Ocal}}$
    & \makecell[l] {Average of a nuclei observable $\wh{\Ocal}$ \\ w.r.t. the nuclei 
                    unperturbed density matrix}
    \\ 
$\displaystyle\vphantom{\int}\bReq=\avg{\whbR(t)}_{\eq}$
    & Equilibrium configuration 
    \\ 
$\displaystyle\vphantom{\int}\bu=\bR-\bReq$
    & Displacement from equilibrium configuration
    \\     
$\displaystyle\vphantom{\int}\wh{\bu}=\whbR-\bReq$
    & Displacement-from-equilibrium-configuration operator 
    \\ 
$\displaystyle\vphantom{\int}\h{\bu}(t)=e^{i\Hnu \hspace {-1pt} t}\h{\bu}\,e^{i\Hnu \hspace {-1pt} t} =\whbR(t)-\bReq$
    & \makecell[l] {Heisenberg evolution of the displacement operator. \\
      It is $\avg{\h{\bu}(t)}_{\eq}=0$}
    \\    
$\displaystyle\vphantom{\int}\rhoeq(\bR)=\Avg{\whrho(\bR,t)}_{\eq}=\braket{\bR|\,\h{\Gamma}{}_{\eq}\,|\bR}$
    & Equilibrium density function 
    \\    
$\displaystyle\vphantom{\int}\delta\whrho(\bR,t)=\whrho(\bR,t)-\rhoeq(\bR)$
    & \makecell[l] {Variation of the density operator w.r.t the equilibrium density.\\
      It is $\avgeq{\delta\whrho(\bR,t)}=0$}
    \\ 
$\displaystyle\vphantom{\int}\delta\Vnu(\bR,t)$
    & Perturbation to the equilibrium nuclei potential $\Vneq(\bR)$
    \\     
$\displaystyle\vphantom{\int}\rho(\bR,t)=
\Tr{\left(\h{\Gamma}{}_{\eq}+\delta\h{\Gamma}(t)\right)\,\whrho(\bR)}
=\rhoeq(\bR)+\braket{\bR|\delta\h{\Gamma}(t)|\bR}$  
    &  \makecell[l] {Expectation value of the nuclei density under the influence of $\delta\Vnu(\bR,t)$,\\
        i.e. using the perturbed density matrix $\h{\Gamma}{}_{\eq}+\delta\h{\Gamma}(t)$}
    \\      
$\displaystyle\vphantom{\int}\delta\rho(\bR,t)=\rho(\bR,t)-\rhoeq(\bR)=\braket{\bR|\delta\h{\Gamma}(t)|\bR}$
    & Time evolution of the variation of the density w.r.t. the equilibrium 
    \\   
\makecell[l] {$\displaystyle\vphantom{\int}\chi(\bR,t,\bR',t')=
     \delta \rho(\bR,t)/\delta \Vnu(\bR',t')$\\
     $\mkern110mu=-i\Theta(t-t')\Avg{[\delta\whrho(\bR,t),\delta\whrho(\bR',t')]}_{\ssst{eq}}$}
    & Nuclei susceptibility
    \\     
$\displaystyle\vphantom{\int}\whVks(\bR)$
    & Kohn-Sham potential operator as a function of the nuclei configuration $\bR$ 
    \\ 
$\displaystyle\vphantom{\int}\displaystyle\vphantom{\int} \hWen(t,t')=\iint d\bR d\bR' \whVks(\bR) \chi(\bR,t,\bR',t')\whVks(\bR')$
    & Effective electron-electron nuclei-mediated interaction operator 
    \\             
$\displaystyle\vphantom{\int}\ket{\bk n}$
    & \makecell[l] {Eigenstate of the equilibrium electronic Hamiltonian\\
      $\hat{H}{}^{\el}=\hat{T}{}^{\el}+\hVks(\bReq)$}
    \\    
$\displaystyle\vphantom{\int}\Vks_{i\bk'\, m\bk}(\bR)=\braket{i\bk'|\hVks(\bR)|m\bk }$
    & \makecell[l]{Braket of the Kohn-Sham potential $\hVks(\bR)$ \\ 
      with the equilibrium electronic eigenstates $\ket{n\bk}$}`
    \\     
$\displaystyle\vphantom{\int}\Velph_{i\bk'\,l\bk\,m\bk\,n\bk'}(t,t')=\braket{i\bk',l\bk|\hWen(t,t')|m\bk ,n\bk'}$   
    & \makecell[l]{Matrix element with the undistorted Bloch eigenstates of the \\
     effective electron-electron nuclei-mediated interaction operator} 
    \\         
$\displaystyle\vphantom{\int}\Sigma(\br,\br';ip_{h})=-\frac{1}{\beta}\sum_l\,G(\br,\br';ip_h-iw_l)\Velph(\br,\br';iw_l)$    
    & \makecell[l]{$G\Wen$ electron self-energy in\\
      spatial coordinate $\br$ and Matsubara frequency $ip_h$ representation} 
    \\           
$\displaystyle\vphantom{\int}\left.\gn^{i\bk\, m\bk'}_{a_1\ldots a_n}=\frac{1}{n!}\frac{\partial^n\Vks_{i\bk\, m\bk'}}{\partial R^{a_1}\ldots \partial R^{a_n}}\right|_{\bReq} $    
    & \makecell[l]{$n-$th order electron-phonon vertex}         
    \\ [1ex]\\
$\displaystyle\vphantom{\int}\gavgn^{i\bk\, m\bk'}_{a_1\ldots a_n}=\frac{1}{n!}\Avgeq{\frac{\partial^n\Vks_{i\bk\, m\bk'}}{\partial R^{a_1}\ldots \partial R^{a_n}}} $    
    & \makecell[l]{Average $n-$th order electron-phonon vertex}         
\end{tabular}
\end{ruledtabular}
\caption{Collection of some of the quantities employed. First column, the symbol used to indicate it, possibly with a definition. Second column, a short description.}
\label{tab:symbols}
\end{table*}
%%%%%%%%%%%%%%%%%%%%%%%%%%%%%%%%%%%%%%%%%%%%%%%%%%%%%%%%%%%%%%%%%%%%%%%%%%%%%%%%%%%%%%%%%%%%%%
%%%%%%%%%%%%%%%%%%%%%%%%%%%%%%%%%%%%%%%%%%%%%%%%%%%%%%%%%%%%%%%%%%%%%%%%%%%%%%%%%%%%%%%%%%%%%%

To have an explicit expression of the nuclei-mediated electron-electron interaction, we need to analyze the three terms in the integrand of Eq.~\eqref{eq:Velph1}.
Nuclei and electrons interact through the Coulomb force, thus if we neglect the electronic screening (and, of course, consider the Coulomb interaction instantaneous), from the probability distribution of the $I$-th nuclei
\begin{equation}
\rho_I(\bR_I,t)=\int\cdots\int\,\rho(\bR,t)\,d\bR_1\ldots d\bR_{I-1}d\bR_{I+1}\ldots d\bR_{N}\,,
\end{equation}
we can write  the interaction energy between electron and the nuclei as
\begin{align}
&-\sum_{I=1}^N\int\frac{Z_I}{|\br-\bR_I|}\,n(\br,t)\rho_I(\bR_I,t)\,d\br d\bR_I \\
&\quad=-\sum_{I=1}^N\int\frac{Z_I}{|\br-\bR_I|}\,n(\br,t)\rho(\bR,t)\,d\br d\bR\,,
\end{align}
where $Z_I$ is the charge of the $I$-th nuclei (we are considering the electronic charge $e=1$). From this 
equation, we readily obtain the bare electron and nuclei potential $\Venot$ and $\Vnnot$, respectively,
\begin{align}
&\Venot(\br,t)=-\sum_{I=1}^N\int\frac{Z_I}{|\br-\bR_I|}\,\rho(\bR,t)\,d\bR\\
&\Vnnot(\bR,t)=-\sum_{I=1}^N\int\frac{Z_I}{|\br-\bR_I|}\,n(\br,t)\,d\br\,,
\end{align}
so that
\begin{equation}
\frac{\delta \Venot(\br,t)}{\delta \rho(\bR,\tau)}=-\sum_{I=1}^N\frac{Z_I}{|\br-\bR_I|}\,\delta(t-\tau)=\frac{\delta \Vnnot(\bR,t)}{\delta n(\br,\tau)}\,.
\end{equation}
Linear-response theory shows that the presence of electronic screening results in replacing $\delta \Venot$ and $\delta \Vnnot$ with 
$\delta \Ve=\varepsilon^{-1} \delta \Venot$ and $\Vn=\varepsilon^{-1} \delta \Vnnot$, respectively,
with $\varepsilon$ the electronic dielectric operator. Therefore, we assume that the same identity applies 
in presence of electronic screening, $\delta \Ve(\br,t)/\delta \rho(\bR,\tau)=\delta \Vn(\bR,t)/\delta n(\br,\tau)$, 
and from the expression of $\delta \Ve(\br,t)/\delta \rho(\bR,\tau)$ coming from Eq.~\eqref{eq:Ve}, we have
\begin{equation}
\frac{\delta \Ve(\br,t)}{\delta \rho(\bR,\tau)}=\Vks(\br,\bR)\,\delta(t-\tau)=\frac{\delta \Vn(\bR,t)}{\delta n(\br,\tau)}\,,
\end{equation}
which insterted into Eq.~\eqref{eq:Velph1} gives
\begin{align}
\Velph(\br,t,\br',t')&=\iint\,d\bR d\bR'\nonumber\\
&\, \Vks(\br,\bR) \frac{\delta \rho(\bR,t)}{\delta \Vn(\bR',t')} \Vks(\br',\bR')\,.
  \label{eq:Velph2}
\end{align}

We are left with the analysis of $\delta \rho(\bR,t)/\delta \Vn(\bR',t')$, which is the nuclear density response to a variation of the nuclei potential. 
This is a crucial contribution, as it is where retardation effects in the nuclei-mediated electron-electron interaction are encoded. From linear-response theory, we have 
\begin{equation}
\delta \rho(\bR,t)=\iint\,d\bR'dt'\,\chi(\bR,t,\bR',t')\delta \Vn(\bR',t')\,,
\end{equation}
where 
\begin{equation}
\chi(\bR,t,\bR',t')=-i\Theta(t-t')\Avg{[\whrho(\bR,t),\whrho(\bR',t')]}_{\ssst{eq}}
\end{equation}
is the nuclear density-density response function (given by a Kubo-type formula), i.e. the lattice susceptibility. In this formula, $\Avg{\,\cdot\,}_{\ssst{eq}}$ is the average in the thermal equilibrium ensemble, and $\hrho(\bR,t)$ is the Heisenberg time evolution ( w.r.t. the unperturbed Hamiltonian) of the density operator $\whrho(\bR)=\delta(\bR-\hbR)$. Subtracting from the density operator the time-constant equilibrium density function, $\delta\whrho(\bR,t)=\whrho(\bR,t)-\rhoeq(\bR)$ , we can also write
\begin{equation}
\chi(\bR,t,\bR',t')=-i\Theta(t-t')\Avg{[\delta\whrho(\bR,t),\delta\whrho(\bR',t')]}_{\ssst{eq}}\,.
\end{equation}
Therefore, we have $\delta \rho(\bR,t)/\delta \Vn(\bR',t')=\chi(\bR,t,\bR',t')$, and substituting this expression into Eq.~\eqref{eq:Velph2} we obtain the final formula
for the nuclei-mediated effective electron-electron interaction potential:
\begin{align}
\Velph(\br,t,\br',t')&=\iint\,d\bR d\bR'\nonumber\\
&\, \Vks(\br,\bR) \chi(\bR,t,\bR',t') \Vks(\br',\bR')\,,
  \label{eq:Velph3}
\end{align}
which written in second quantization in the Heisenberg representation is 
\begin{align}
\Velphhat(t,t')&=\iint\,d\br d\br'\,\Velph(\br,t,\br',t')\nonumber \\
&\qquad\,\whpsi(\br,t){}^{\dagger}\whpsi{}^{\dagger}(\br',t')\whpsi(\br',t')\whpsi(\br,t)\,.
\end{align}
Notice that if it were something like $\Velph(\br,t,\br',t')=V(\br,\br')\delta(t-t')$, with $V(\br,\br')$ a generic 2-point function, 
we would have an instantaneous two-body interaction (like the Coulomb interaction in non-relativistic regime, where $V(\br,\br')\propto 1/|\br-\br'|$),
whereas now with the non-trivial time-dependence encoded in $\Velph(\br,t,\br',t')$ we have the retardation effects of the nuclei-mediated interaction.
%It is interesting to observe that from Eq.~\eqref{eq:Ve} we have that the variation of the electron potential  $\delta\Ve(\br,t)$ is given by
%\begin{equation}
%\delta\Ve(\br,t)=\int\,\Vks(\br,\bR)\,\delta\rho(\bR,t)\,d\bR\,,
%\end{equation}
%thus we can write the nuclei-mediated effective electron-electron interaction in the perspicuous form
%\begin{equation}
%\Velph(\br,t,\br',t')=-i\Theta(t-t')\Avg{[\delta\Ve(\br,t),\delta\Ve(\br',t')]}_{\ssst{eq}}\,.
%\end{equation}
Using $\Velphhat(t,t')$, we can use all the machinery of the theoretic-field approach to fully include the EPI effects in the electron dynamics, defining the electron self-energy due to the electron-nuclei coupling $\Sigma^{\elph}$, and producing a Dyson-like equation for the electron Green function. 

Before doing that, we analyze more in detail $\Velphhat(t,t')$. For the following derivations, it is convenient to use the representation of $\Velphhat(t,t')$ in the Bloch eigenstates of $\Hel$, the electronic Hamiltonian with ions frozen at equilibrium 
(i.e. the Kohn-Sham Hamiltonian with potential $\VKS(\br,\bReq)$). Therefore, we consider a (macroscopic) supercell made of $N_c$ unit cells, with periodic boundary conditions (PBCs), 
and the discrete orthonormal basis set of eigenstates $\Ket{{n \bk }}$ with energy $\epsilon_{n \bk}$,
where $n$ is the band index, and $\bk$ is the pseudo-momentum, belonging to the Brillouin-zone (BZ) and satisfying the supercell Born–von K\'arm\'an (BvK) condition:
\begin{align}
&\Braket{\bk n | \bk' n'}=\delta_{\bk \bk'}\delta_{n n'}\\
&\Hel\Ket{{n \bk }}=\epsilon_{n \bk}\Ket{{n \bk }}\,.
\end{align}
In coordinates representation, these eigenstates are
\begin{equation*}
\Braket{\br|\psi_{n\bk}}=\psi_{n\bk}(\br)=\frac{1}{\sqrt{\Nc}}u_{n\bk}(\br)e^{i\bk\cdot\br}\,,
\end{equation*}
where $u_{n\bk}(\br)$ are lattice-periodic functions normalized to one in the crystal unit cell.
Using the basis set $\Ket{{n \bk }}$, and the corresponding creation and annihilation operators $\hat{c}{}^{\dagger}_{n\bk}$ and $\hat{c}_{n\bk}$, respectively, we can write
\begin{align}
\Velphhat(t,t')&=\sum_{\bk,\bk',ilmn}\Velph_{i\bk'\,l\bk\,m\bk\,n\bk'}(t,t')\nonumber\\
&\quad \hat{c}{}^{\dagger}_{i \bk' }(t)\hat{c}{}^{\dagger}_{l \bk }(t')\hat{c}{}_{n \bk' }(t')\hat{c}{}_{ m \bk}(t)\,,
\end{align}
with
\begin{align}
\Velph_{i\bk'\, l\bk\, m \bk\, n\bk'}(t,t')&=\Braket{i\bk',l\bk|\Velphhat(t,t')|m\bk ,n\bk'}\\
&\mkern-150mu =\iint\,d\br d\br' \Velph(\br,t,\br',t')\nonumber \\
&\mkern-10mu \psi^*_{i\bk'}(\br)\psi^*_{l\bk}(\br')\psi_{n\bk'}(\br')\psi_{m\bk}(\br)\\
&\mkern-150mu =\iint\,d\bR d\bR' \,\Vks_{i\bk'\, m\bk }(\bR)\chi(\bR,t,\bR',t')\Vks_{l\bk \,n\bk' }(\bR')\,,
\end{align}
where 
\begin{align}
\Vks_{i\bk'\, m\bk}(\bR)&=\Braket{i\bk'|\Vkshat(\bR)|m\bk }\nonumber\\
&=\!\!\int d\br\,\psi^*_{i\bk'}(\br)\,\Vks(\br,\bR)\,\psi_{m\bk}(\br)\,,
\end{align}
and the integrations on $\br$ are intended to be over the PBCs supercell.

\begin{figure}[t!]
\includegraphics[width=\columnwidth]{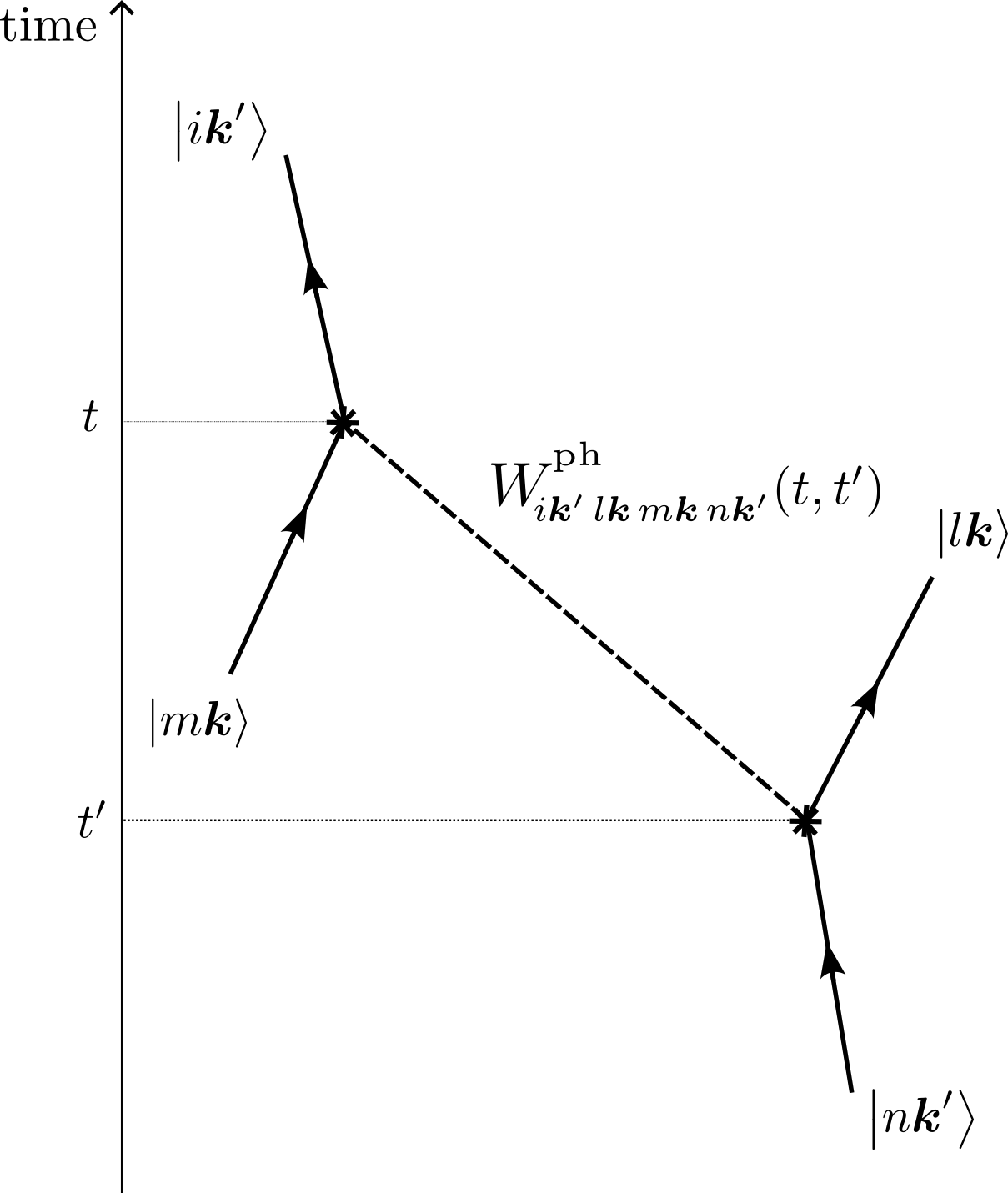}
\caption{Diagrammatic representation of the nuclei-mediated effective electron-electron interaction between Bloch states. Due to this interaction, at time $t'$  an electron scatters from $\Ket{n\bk'}$ to $\Ket{l\bk}$, which causes  the scattering of an electron from $\Ket{m\bk}$ to $\Ket{i\bk'}$ at a later time $t$.}
\label{fig:elph_scattering}
\end{figure}

In Fig.~\ref{fig:elph_scattering} we have the diagrammatic representation of this interaction. It is worthwhile to stress that
the crystalline description in terms of unit cell and PBCs supercell refers to the undistorted atomic configuration $\bReq$.
The distorted atomic configurations that we are considering here, $\bR=\bReq+\bu$, are totally general and have only the periodicity 
of the PBCs supercell. That is why the nuclei-mediated electron-electron interaction can couple any momenta $\bk$ and $\bk'$. Of course, we could decompose the atomic distortions in wave-modulated displacements with a certain wave vector $\bq$, so as to make independent analysis for different monochromatic distortions. However, in order to keep the description as simple as possible, we will avoid this further complication. Nevertheless, notice that we have already employed the momentum conservation (i.e. total momentum in the interaction is conserved), which is due to system's translational invariance, and corresponds to the fact that in $\Velph(\br,t,\br',t')$ only the difference $\br'-\br$ is significant. 
Similarly, since we have time-translation symmetry, the notation can be simplified further. In all these quantities only time differences are significant, so that we can employ quantities like  $\chi(\bR,\bR',t)$ and $\Velph(\br,\br',t')$, which are defined by
$\chi(\bR,\bR',t-t')=\chi(\bR,t,\bR',t')$ and $\Velph(\br,\br',t-t')=\Velph(\br,t,\br',t')$, respectively.  Therefore, we can write:
\begin{align}
&\Velph_{i\bk'\, l\bk\, m \bk\, n\bk'}(t)=\nonumber\\
&\mkern50mu \iint\,d\bR d\bR'\Vks_{i\bk'\, m\bk } \chi(\bR,\bR',t) \Vks_{l\bk \,n\bk' }(\bR').
\label{eq:Velphtau}
\end{align}
We could Fourier transform these quantities w.r.t the time and obtain the description in terms of energy. However, in order to include temperature effects, we employ the imaginary-time formalism and Fourier transform these quantities to the ``Matsubara frequencies'' representation. In this way we obtain ($\hbar=1$):
\begin{align}
&\Velph_{i\bk'\, l\bk\, m \bk\, n\bk'}(i\omega_\nu)=\nonumber\\
&\mkern50mu \iint\,d\bR d\bR'\Vks_{i\bk'\, m\bk }(\bR)  \chi(\bR,\bR',i\omega_\nu) \Vks_{l\bk \,n\bk' }(\bR')
\label{eq:Velphmat}\\
&\chi(\bR,\bR',i\omega_\nu)=\int_0^\beta d\tau\,e^{i\omega_\nu\tau}\chi(\bR,\bR',\tau)\\
&\chi(\bR,\bR',\tau)=-\Avg{T_{\tau}\delta\whrho(\bR,\tau)\delta\whrho(\bR',0)}_{\ssst{eq}}\,.
\end{align}
where $\omega_\nu=2\pi\nu/\beta$, with $\nu\in\mathbb{Z}$ and $\beta=1/\kB T$, with $k_B$ Boltzmann's constant.
%%%%%%%%%%%%%%%%%%%%%%%%%%%%%%%%%%%%%%%%%%%%%%%%%%%%%%%%%%%%%%%%%%%%

%%%%%%%%%%%%%%%%%%%%%%%%%%%%%%%%%%%%%%%%%%%%%%%%%%%
\section{The linear order}
\label{sec:the_linear_order}
\begin{figure}[t]
\includegraphics[width=\columnwidth]{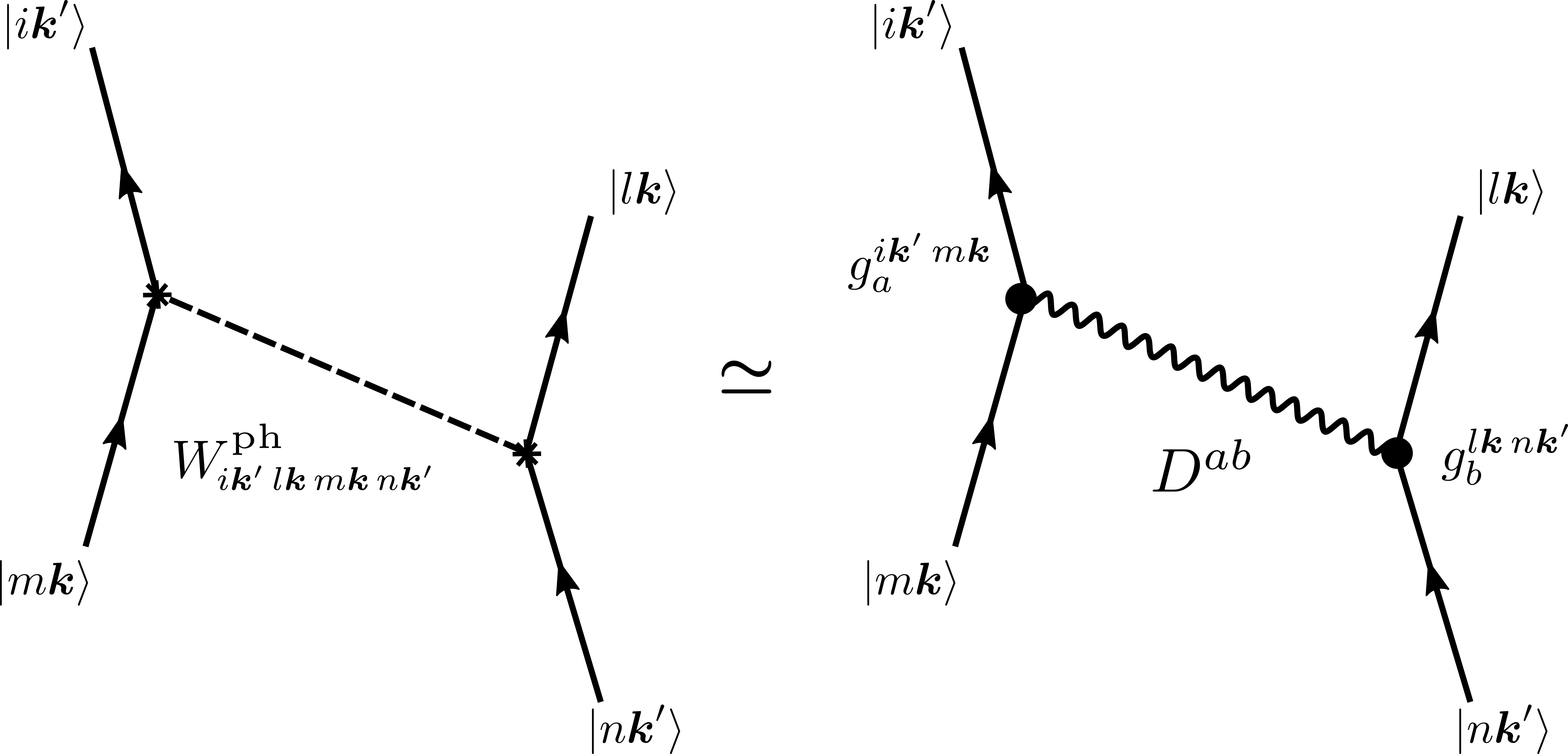}
\caption{Diagrammatic representation of the nuclei-mediated effective electron-electron interaction between Bloch states at lowest order in the atomic displacement from the equilibrium configuration.}
\label{fig:Velph_lin}
\end{figure}
The derivation done in the previous section is general. Indeed, it applies even when the atomic displacements are large.  Here we consider the lowest-order in the atomic displacements from the equilibrium configuration to verify that we obtain the standard approach. At equilibrium, the density $\rhoeq(\bR)$ is time independent. The equilibrium atomic configuration is, by definition, the average atomic configuration according to $\rhoeq(\bR)$, $\bReq=\Avg{\bR}_{\eq}$, and therefore the displacement from the equilibrium configuration $\hbu=\hbR-\bReq$ has average zero at equilibrium $\Avg{\hbu}_{\eq}=0$.
The probability density operator is
\begin{align}
\hat{\rho}(\bR)&=\delta(\bR-\hat{\bR})\vphantom{\sum_a\hat{u}}\\
&=\delta(\bR-\bReq-\hat{\bu})\vphantom{\sum_a\hat{u}}\\
&=\delta(\bR-\bReq)-\sum_a\hat{u}{}^a\partial_a\delta(\bR-\bReq)\nonumber\\
&\mkern80mu +\frac{1}{2}\sum_{ab}\hat{u}{}^a\hat{u}{}^b\partial^2_{ab}\delta(\bR-\bReq)+\ldots\,,
\end{align}
where $a=(\alpha,s)$ and $b=(\beta,t)$ are composite Cartesian $\alpha,\beta$ and atomic $s,t$ indices.
From the previous equation, since $\Avg{\hbu}_{\eq}=0$, we have
\begin{align}
\delta\hat{\rho}(\bR)&=\hat{\rho}(\bR)-\Avg{\hat{\rho}(\bR)}_{\eq}\\
&=-\sum_a\hat{u}{}^a\partial_a\delta(\bR-\bReq)+\Ocal(u^2)\,
\end{align}
and therefore
\begin{align}
&\chi(\bR,\bR',\tau)=-\Avg{T_{\tau}\delta\hrho(\bR,\tau)\delta\hrho(\bR',0)}_{\eq}\\
&\quad=\sum_{ab}D^{ab}(\tau)\partial_a\delta(\bR-\bR_{\eq})\partial_b\delta(\bR'-\bR_{\eq})+\Ocal(u^3)\,,
\label{eq:chi_lowest}
\end{align}
where
\begin{equation}
D^{ab}(\tau)=-\Avg{T_{\tau}\hu{}^a(\tau)\hu{}^b(0)}_{\eq}
\end{equation}
is the atomic displacement-displacement correlation function (i.e. the full phonon Green function, in general including anharmonic effects, since we are not just considering harmonic dynamics for nuclei).
Plugging Eq.~\eqref{eq:chi_lowest} in Eq.~\eqref{eq:Velph3} and using the properties of the Dirac delta we obtain
\begin{align}
&\Velph(\br,\br',\tau)\nonumber\\
&=\sum_{ab}D^{ab}(\tau)\iint\,d\bR d\bR'\Vks(\br,\bR) \partial_a\delta(\bR-\bR_{\eq})\nonumber\\
&\mkern80mu \times\partial_b\delta(\bR'-\bR_{\eq}) \Vks(\br',\bR')+\Ocal(u^3)\\
&=\sum_{ab}\partial_a\Vks(\br,\bReq)\,D^{ab}(\tau)\,\partial_b\Vks(\br',\bReq)+\Ocal(u^3)
\label{eq:VelphLow}
\end{align}
and using the Bloch eigenstates
\begin{equation}
\Velph_{i\bk'\, l\bk\, m \bk\, n\bk'}(\tau)=\sum_{ab}g^{i\bk'\,m\bk}_aD^{ab}(\tau)g^{l\bk\,n\bk'}_b+\Ocal(u^3)\,,
\label{eq:VelphBlochLow}
\end{equation}
where $g^{i\bk'\,m\bk}_a$ is the standard first-order (i.e. linear) electron-phonon matrix element
\begin{equation}
g^{i\bk'\,m\bk}_a=\Braket{i\bk'|\left.\frac{\partial \whVks}{\partial R^a}\right|_{\bReq}|m\bk}=\left.\partial_a\Vks_{i\bk'\,m\bk}\right|_{\bReq}\,.
\end{equation}
At the lowest order in the displacement $u$, Eq.~\eqref{eq:VelphBlochLow}  returns the standard formula used to evaluate the EPI, and it is diagrammatically  described in Fig.~\ref{fig:Velph_lin}. As said, we have not make any assumption about the ionic potential.  If, somehow consistently with the fact that  we are considering the lowest order in the atomic displacement, we consider the harmonic approximation, the nuclei Green function becomes the non-interacting phonon Green function, and we fully recover the standard approach. In this case, since the  equilibrium density matrix is normal w.r.t the displacement, then $\Avg{u^{2n+1}}_{\eq}=0$ (i.e. the average of an odd power of the displacements is zero) and thus Eqs.~\eqref{eq:chi_lowest},~\eqref{eq:VelphLow}, and~\eqref{eq:VelphBlochLow} are correct up to the 4-th order in $u$. 

It is worthwhile to conclude this section with an important remark. The statement ``lowest order in the atomic displacement'' is quite vague, and we need to be more precise. A closer look at Eq.~\eqref{eq:VelphLow} makes clear that what is relevant is the variation of the Kohn-Sham potential with respect to the atomic displacement (typically, as it will be clearer later, the matrix element $\Vks_{i\bk'\,m\bk}$ with states close to the Fermi level). In other words,  truncating the nuclei-mediated electron-electron interaction at the lowest order in the atomic displacement can be considered a fair approximation as long as the variation of the matrix element of the Kohn-Sham potential between states of interest can be neglected for the typical displacements the atoms undergo. On the other side, the harmonic approximation is appropriate as long as the variation of the atomic potential w.r.t the atomic displacements can be neglected beyond the second order. Therefore, it must be kept in mind that these two criteria, although somehow correlated, in principle are different, one considering the intra-nuclei potential and the other one the electron-nuclei potential.
%%%%%%%%%%%%%%%%%%%%%%%%%%%%%%%%%%%%%%%%%%%%%%%%%%%%%%%%%%%%%%%%%%%%

%%%%%%%%%%%%%%%%%%%%%%%%%%%%%%%%%%%%%%%%%%%%%%%%%%%%%%%%%%%%%%%%%%%%
\section{Nuclei with Gaussian space distribution}
\label{sec:nuclei_with_harmonic_dynamics}
In Sec.~\ref{sec:The_nuclei-mediated_electron-electron_interaction} we presented the general expression of the nuclei mediated electron-electron effective interaction, which is applicable for the most generic nuclei dynamics.  Now we consider, in particular, the case where the nuclei equilibrium distribution function $\rhoeq(\bu)$ is a (multivariate) Gaussian distribution. This is the case, for instance, when the nuclei potential is quadratic (i.e. harmonic). Of course, this does not necessarily mean that we are considering the standard harmonic approximation, i.e. that we are truncating the Taylor expansion of the ionic potential-energy surface (PES) around its minimum at the second order. As a matter of fact, we can consider, for example, the effective harmonic potential obtained within the self-consistent harmonic approximation, i.e. the quadratic potential that minimizes the ionic free-energy functional, and the corresponding Gaussian distribution function $\rhoeq(\bu)$. Indeed, the stochastic implementation of this method, the so-called SSCHA, has been extensively used to include the anharmonic (and quantum) effects of lattice dynamics at a non-perturbative level in several cases~\cite{Monacelli_2021}. Therefore, consistently with the scope of this work, we can include the case of large atomic displacements, which is beyond the harmonic approximation regime, and still use a Gaussian nuclei distribution function. As we will see, this allows us to exploit some special results that yield to an analytic expression for $\chi(\bR,\bR',\tau)$. 

Given two generic operators $\hA$ and $\hB$ it is
\begin{equation}
\Avg{\left[\hA-\Avg{\hA}\right]\left[\hB-\Avg{\hB}\right]}=\Avg{\hA\hB}-\Avg{\hA}\Avg{\hB}.
\end{equation}
Thus, from $\hat{\rho}(\bR)=\delta(\bR-\hat{\bR})=\delta(\bR-\bReq-\hat{\bu})$, for $\tau>0$ we have
\begin{align}
\chi(\bR,\bR',\tau)&=\Avg{\delta(\bR-\bReq-\hat{\bu}(\tau))}_{\eq}\Avg{\delta(\bR'-\bReq-\hat{\bu}(0))}_{\eq}\nonumber\\
&\quad-\Avg{\delta(\bR-\bReq-\hat{\bu}(\tau))\delta(\bR'-\bReq-\hat{\bu}(0))}_{\eq}\,,
\label{eq:chi_2}
\end{align}
and using the Fourier transform
\begin{equation}
\delta(\bR-\bReq-\hbu)=\frac{1}{(2\pi)^{3N}}\int\,d\bq\,e^{-i\bq\cdot(\bR-\bReq-\hbu)}\, ,
\end{equation}
we have
\begin{widetext}
\begin{align}
\chi(\bR,\bR',\tau)&=\frac{1}{(2\pi)^{6N}}\iint\,d\bq d\bq'
\left[\Avg{e^{-i\bq\cdot(\bR-\bReq-\hat{\bu}(\tau))}}_{\eq}\Avg{e^{-i\bq'\cdot(\bR'-\bReq-\hat{\bu}(0))}}_{\eq}
-\Avg{e^{-i\bq\cdot(\bR-\bReq-\hat{\bu}(\tau))}e^{-i\bq'\cdot(\bR'-\bReq-\hat{\bu}(0))}}_{\eq}\right]\\
&=\frac{1}{(2\pi)^{6N}}\iint\,d\bq d\bq'e^{-i\bq\cdot(\bR-\bReq)}e^{-i\bq'\cdot(\bR'-\bReq)}
\left[\Avgeq{e^{-i\bq\cdot\hat{\bu}(\tau)}}\Avgeq{e^{-i\bq'\cdot\hat{\bu}(0)}}-\Avgeq{e^{-i\bq\cdot\hat{\bu}(\tau)}e^{-i\bq'\cdot\hat{\bu}(0)}}\right]\,.
\label{eq:chi_3_taupos}
\end{align}
\end{widetext}
A similar formula is obtained for $\tau<0$, by changing the order of the operators, consistently with the time-ordering operator $T_{\tau}$.

Now we consider a normal equilibrium distribution function for the nuclei, i.e. the equilibrium distribution function for the nuclei interacting through the quadratic potential
$\sum_{ab}u^a\phi_{ab}u^b$, where $\phi_{ab}$ is the so-called force-constants matrix FC. It is
\begin{equation}
\rhoeq(\bR)=\frac{1}{\sqrt{\textup{det}(2\pi\bPsi)}}e^{-\frac{1}{2}\bu\cdot\bPsi^{-1}\cdot\bu}\,,
\label{eq:rhoeqharm}
\end{equation}
with
\begin{equation}
\Psi^{ab}=\Avgeq{\hu{}^a \hu{}^b}=\sum_{\mu}\frac{1+2n_\mu}{2\omega_\mu}\frac{e^a_{\mu}}{\sqrt{M_a}}\frac{e^b_{\mu}}{\sqrt{M_b}}\,
\label{eq:Psi_def}
\end{equation}
the displacement-displacement correlation matrix, where $M_a$ is the atomic mass of atom $a$, $\omega_{\mu}$ and $e^a_{\mu}$ are phonon frequencies and modes of the harmonic potential, respectively  (i.e. $\omega_{\mu}^2$ and $e^a_{\mu}$ are eigenvalues and eigenvectors of the dynamical matrix $D_{ab}=\phi_{ab}/\sqrt{M_a M_b}$, respectively), and $n_\mu=1/(e^{\beta\omega_{\mu}}-1)$ is the Bose occupation factor of the mode $\mu$.

Given two any operators $\hA$ and $\hB$ which are linear in the displacements (and momenta), the thermal equilibrium average w.r.t a normal distribution fulfils this relation~\cite{ashcroft1976solid, doi}
\begin{equation}
\Avgeq{e^{\hA}e^{\hB}}=e^{\frac{1}{2}\Avgeq{\hA{}^2+2\hA\hB+\hB{}^2}}\,,
\end{equation}
therefore
\begin{align}
&\Avgeq{e^{\hA}}\Avgeq{e^{\hB}}-\Avgeq{e^{\hA}e^{\hB}}=e^{\frac{1}{2}\Avgeq{{\hA}{}^2}}\,\, e^{\frac{1}{2}\Avgeq{\hB{}^2}}
\nonumber\\
&\mkern300mu\times\left[1-e^{\Avgeq{\hA\hB}}\right]\,.
\label{eq:step1}
\end{align}
It is
\begin{equation}
\Avgeq{(\bq\cdot\hbu(\tau))^2}=\Avgeq{(\bq\cdot\hbu(0))^2}=\Avgeq{(\bq\cdot\hbu)^2}=\bq\cdot\bPsi\cdot\bq\,,
\label{eq:step2}
\end{equation}
and
\begin{equation}
\Avgeq{T_{\tau}(\bq\cdot\hbu(\tau))(\bq'\cdot\hbu(0))}=-\bq\cdot\bD(\tau)\cdot\bq'\,,
\label{eq:step3}
\end{equation}
where $D^{ab}(\tau)=-\Avgeq{T_{\tau}\hu{}^a(\tau)\hu{}^b(0)}$
is the harmonic (i.e. non-interacting) phonon Green function.
Using Eqs.~\eqref{eq:chi_3_taupos},~\eqref{eq:step1},~\eqref{eq:step2},~\eqref{eq:step3}, and making the changes of variables $\bR(\bR')\rightarrow\bReq+\bu(\bu')$ in the integrals of Eq.~\eqref{eq:Velphtau}, we can write
\begin{align}
&\Velph(\br,\br',\tau)=\nonumber\\
&\mkern50mu \iint\,d\bu d\bu'\Vks(\br,\bReq+\bu) \chi(\bu,\bu',\tau) \Vks(\br',\bReq+\bu')
\end{align}
and, therefore,
\begin{align}
&\Velph_{i\bk'\, l\bk\, m \bk\, n\bk'}(\tau)=\nonumber\\
&\mkern50mu \iint\,d\bu d\bu'
\Vks_{i\bk'\, m\bk }(\bReq+\bu) \chi(\bu,\bu',\tau) \Vks_{l\bk \,n\bk' }(\bReq+\bu')\,,
\label{eq:Velph_harm}
\end{align}
with
\begin{align}
&\chi(\bu,\bu',\tau)=\nonumber\\
&\mkern40mu\frac{1}{(2\pi)^{2N}}\iint d\bq d\bq'e^{-i\bq\cdot\bu}e^{-i\bq'\cdot\bu'}\nonumber\\
&\mkern120mu e^{-\frac{1}{2}\bq\cdot\bPsi\bq}e^{-\frac{1}{2}\bq'\cdot\bPsi\bq'}\left[1-e^{\bq\cdot\bD(\tau)\cdot\bq'}\right]
\label{eq:chi_gauss}\,.
\end{align}

Eqs. \eqref{eq:Velph_harm} and \eqref{eq:chi_gauss} are the general expressions of the nuclei-mediated electron-electron interaction when the nuclei have a Gaussian space distribution. In principle, these equations allow the numerical computation of the nuclei-mediated electron-electron interaction, creating at a given complex time $\tau$ ionic configurations that obey Eq.~\eqref{eq:chi_gauss}, and computing the integral Eq.~\eqref{eq:Velph_harm}. However, adopting this direct computational approach would pose technical difficulties, due to the presence of $\tau$ and the need of analytical continuation techniques. For this reason, in the next section we will analyze further the obtained results, by performing a Taylor series analysis and describing the obtained terms in a diagrammatic fashion. The goal of this analysis is twofold. On the one hand, we will establish the connection with the standard electron-phonon perturbative analysis, making explicit what are the contributions included in the presented nuclei-mediated electron-electron effective interaction. On the other hand, we will show a computational procedure  that can be easily implemented in conjunction with \emph{ab initio} calculations, and
that allows to obtain results whose accuracy can be systematically improved. 
%%%%%%%%%%%%%%%%%%%%%%%%%%%%%%%%%%%%%%%%%%%%%%%%%%%%%%%%%%%%%%%%%%%%%%%%%%%%%%%%%%%%%%%%%

%%%%%%%%%%%%%%%%%%%%%%%%%%%%%%%%%%%%%%%%%%%%%%%%%%%%%%%%%%%%%%%%%%%%%%%%%%%%%%%%%%%%%%%%%
\section{Nuclei with Gaussian space distribution: a diagrammatic analysis}
\label{sec:nuclei_with_harmonic_dynamics_PDA}
\begin{figure*}[t!]
\includegraphics[width=\textwidth]{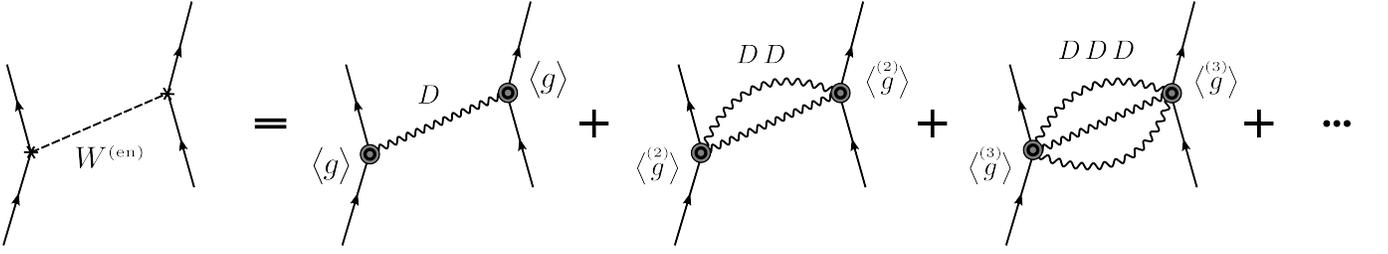}
\caption{Diagrammatic expansion of the nuclei-mediated electron-electron effective interaction $\Velph$ as described in Eqs.~\eqref{eq:Velph_Gauss_freq_series},~\eqref{eq:Velph_Gauss_freq_nth},  and~\eqref{eq:convol_phon}.}
\label{fig:multiphonon}
\end{figure*}

Doing a Taylor expansion of the exponential in Eq.~\eqref{eq:chi_gauss}, we obtain
\begin{align}
&\chi(\bu,\bu',\tau)\nonumber\\
 &\,\,=-\sum_{n=1}^{+\infty}\frac{1}{n!}\,
 \sum_{\substack{a_1\cdots a_n \\ b_1\cdots b_n}}
 \Lambda_{a_1\ldots a_n}(\bu)
 \left[\prod_{h=1}^{n}D^{a_h b_h}(\tau)\right]
 \Lambda_{b_1\ldots b_n}(\bu')\,,
\label{eq:chitaylor}
\end{align}
with
\begin{align}
&\Lambda_{a_1\ldots a_n}(\bu)=\frac{1}{(2\pi)^N}\int d\bq\,\left({q}^{a_1} \ldots {q}^{a_n}\right)\, e^{-i\bq\cdot\bu}e^{-\frac{1}{2}\bq\cdot\bPsi\bq}\\
&=i^n\frac{\partial^n}{\partial {u}^{a_1}\ldots \partial {u}^{a_n}}\frac{1}{(2\pi)^N}\int d\bq\,e^{-i\bq\cdot\bu}e^{-\frac{1}{2}\bq\cdot\bPsi\bq}\\
&=i^n\frac{\partial^n}{\partial {u}^{a_1}\ldots \partial {u}^{a_n}}\frac{1}{\sqrt{\text{det}(2\pi\bPsi)}}e^{-\frac{1}{2}\bu\cdot\bPsi^{-1}\bu}\\
&=i^n\frac{\partial^n \rhoeq}{\partial {u}^{a_1}\ldots \partial {u}^{a_n}}\,,
\label{eq:Lambda}
\end{align}
where we have used the well-known properties of the Fourier transform of $n$-dimensional Gaussian functions, and the expression Eq.~\eqref{eq:rhoeqharm} of the harmonic equilibrium distribution function $\rhoeq(\bu)$. Inserting Eqs.~\eqref{eq:chitaylor} and~\eqref{eq:Lambda} in Eq.~\eqref{eq:Velph_harm}, and using integration by parts in terms like
\begin{align}
&\int\,d\bu\,\Vks_{i\bk'\,m\bk}(\bReq+\bu)\frac{\partial^n \rhoeq}{\partial {u}^{a_1}\ldots \partial {u}^{a_n}}\nonumber\\
&\mkern80mu =(-1)^n\int\,d\bu\left.\frac{\partial^n \Vks_{i\bk'\,m\bk}}{\partial {u}^{a_1}\ldots \partial {u}^{a_n}}\right|_{\bReq+\bu}\rhoeq(\bu)\\
&\mkern80mu =(-1)^n\Avgeq{\frac{\partial^n \Vks_{i\bk'\,m\bk}}{\partial {R}^{a_1}\ldots \partial {R}^{a_n}}}\,,
\end{align}
we can expand the nuclei-mediated interaction as a series
\begin{equation}
\Velph_{i\bk'\, l\bk\, m \bk\, n\bk'}(\tau)=
\sum_{n=1}^{+\infty}\,\Wenn_{i\bk'\, l\bk\, m \bk\, n\bk'}(\tau)
\label{eq:Velph_Gauss_tau_series}
\end{equation}
with the $n$-th term given by
\begin{align}
&\Wenn_{i\bk'\, l\bk\, m \bk\, n\bk'}(\tau)=\nonumber \\
&\quad n!\sum_{\substack{a_1\cdots a_n \\ b_1\cdots b_n}}\gavgn^{i\bk'\, m\bk}_{a_1\ldots a_n}
\left[(-1)^{n-1}\prod_{h=1}^{n}D^{a_h b_h}(\tau)\right]
\gavgn^{l\bk \, n\bk'}_{b_1\ldots b_n} ,
\label{eq:Velph_Gauss_tau_nth}
\end{align}
where we have introduced the average $n$-th order vertex %$\gn$
\begin{equation}
\gavgn^{i\bk'\, m\bk}_{a_1\ldots a_n}=\frac{1}{n!}\Avgeq{\frac{\partial^n\Vks_{i\bk'\, m\bk}}{\partial R^{a_1}\ldots \partial R^{a_n}}} 
\label{eq:def_gavgn}
\end{equation}
This can be compared with the standard $n$-th order bare electron-phonon vertex
\begin{equation}
\gn^{i\bk'\, m\bk}_{a_1\cdots a_n}=\left.\frac{1}{n!}\,\frac{\partial^n\Vks_{i\bk'\, m\bk}}{\partial R^{a_1}\ldots \partial R^{a_n}}\right|_{\bReq}\,.
\label{eq:def_gn}
\end{equation}
The interpretation of Eq.~\eqref{eq:Velph_Gauss_tau_nth} is especially clear if we consider the Fourier transform to the Matsubara frequencies: 
\begin{equation}
\Wenn_{i\bk'\, l\bk\, m \bk\, n\bk'}(i\omega_l)=
\sum_{n=1}^{+\infty}\,\Wenn_{i\bk'\, l\bk\, m \bk\, n\bk'}(i\omega_l)
\label{eq:Velph_Gauss_freq_series}
\end{equation}
with
\begin{align}
&\Wenn_{i\bk'\, l\bk\, m \bk\, n\bk'}(i\omega_l)=\nonumber\\
&\quad n!\sum_{\substack{a_1\cdots a_n \\ b_1\cdots b_n}}
\gavgn^{i\bk'\, m\bk}_{a_1\ldots a_n}
\left[(-1)^{n-1}\left(\Conv_{h=1}^{n}\Bcal^{a_h b_h}\right)(i\omega_l)\right]
\gavgn^{l\bk \, n\bk'}_{b_1\ldots b_n}\,,
\label{eq:Velph_Gauss_freq_nth}
\end{align}
where 
\begin{align}
\Bcal^{a b}(i\omega_l)&=\int_0^{\beta}\,d\tau\,e^{i\omega_l\tau}D^{ab}(\tau)\\
&=\Bigl[(i\omega_l)^2M_a\delta_{ab}-\phi_{ab}\Bigr]^{-1}
\end{align}
is the Fourier transform of the phonon Green function, and
\begin{equation}
\left(\Conv_{h=1}^{n}\Bcal^{a_h b_h}\right)(i\omega_l)=\int_0^{\beta}\,d\tau\,e^{i\omega_l\tau}\left[\prod_{h=1}^{n}D^{a_h b_h}(\tau)\right]\\
\end{equation}
is the convolution of the Fourier transform of $n$ phonon Green functions, coming from the Fourier transform of the product of $n$ phonon Green functions. It is
\begin{align}
&(-1)^{n-1}\left(\,\Conv_{h=1}^{n}\Bcal^{a_h b_h}\,\right)(i\omega_l)\nonumber\\
&\vphantom{\frac{1}{\beta}}\mkern20mu =(-1)^{n-1}\left(\Bcal^{a_1 b_1}\circledast \Bcal^{a_2 b_2}\circledast \ldots\circledast \Bcal^{a_n b_n}\right)(i\omega_l)\nonumber\\
&\mkern20mu =\left(-\frac{1}{\beta}\right)\sum_{m_1}\Bcal^{a_1 b_1}(i\omega_{m_1})\times
            \cdots\nonumber\\
&\mkern70mu       \cdots\times      \left(-\frac{1}{\beta}\right)\sum_{m_{n-1}}\Bcal^{a_{n-1}b_{n-1}}(i\omega_{m_{n-1}})\nonumber\\
&\vphantom{\frac{1}{\beta}}\mkern130mu\times\Bcal^{a_n b_n}(i\omega_l-i\omega_{m_1}-\ldots-i\omega_{m_{n-1}})\,.
\label{eq:convol_phon}
\end{align}
From Eqs.~\eqref{eq:Velph_Gauss_freq_series},~\eqref{eq:Velph_Gauss_freq_nth},  and~\eqref{eq:convol_phon} we see that the nuclei-mediated electron-electron effective interaction is given by the sum of infinite contributions, the $n$-th order contribution corresponding to $n$ phonon propagators connecting two average $n$-th order vertices $\gavgn$ ($n!$ being just a combinatorial factor to take into account the different way of pairing them). In Fig.~\ref{fig:multiphonon} we show the diagrams 
describing this expansion.

\begin{figure}[t!]
\includegraphics[width=\columnwidth]{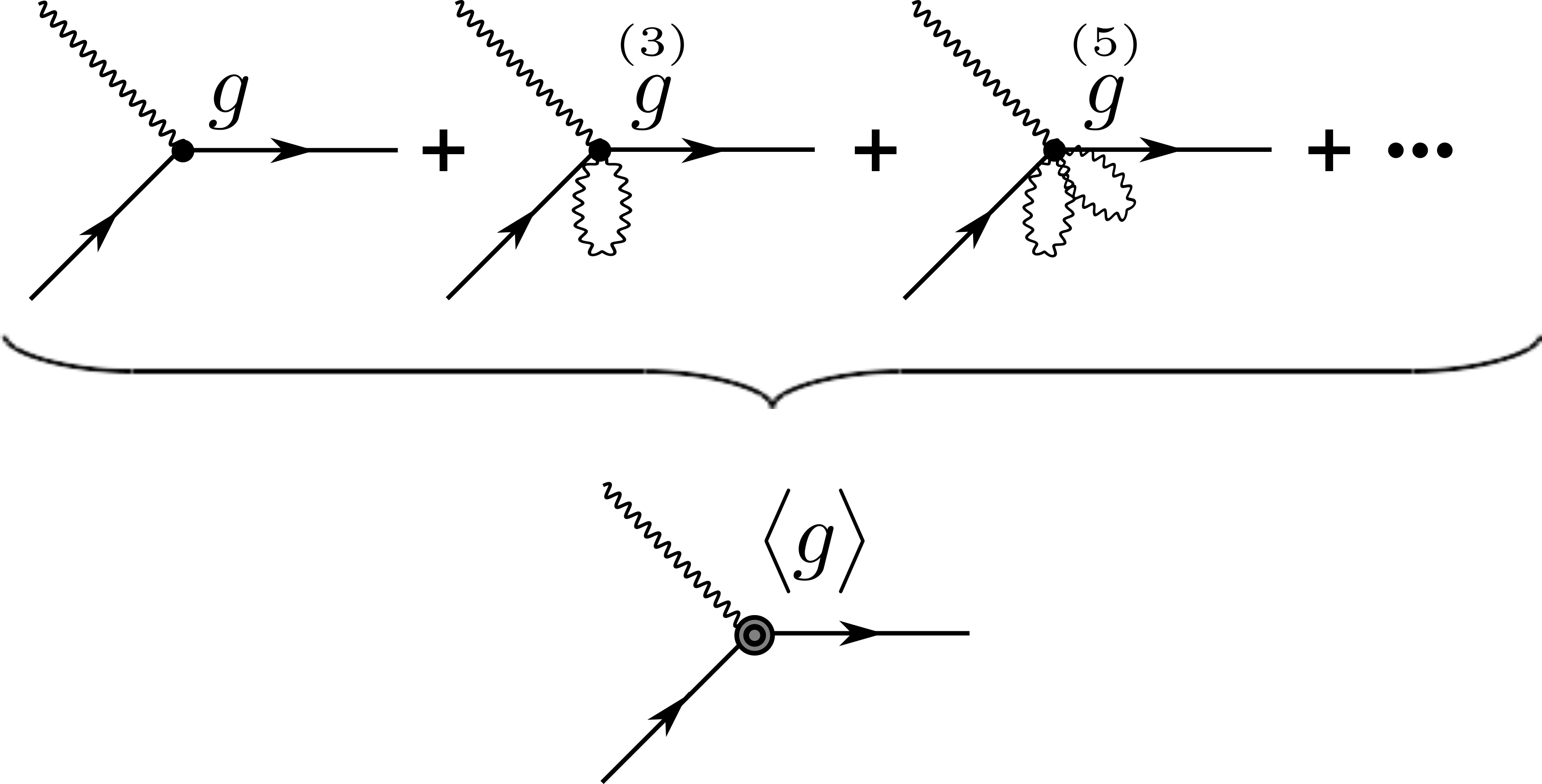}
\caption{Average single electron-phonon vertex $\gavg$, given by the superposition of simple $g$, triple $\gx{3}$, quintuple $\gx{5}$ \ldots 
phonon vertices with only one external phonon leg, and the other phonon legs paired in loops.}
\label{fig:gavg}
\end{figure}

We are left with the interpretation of the average electron-phonon vertices $\gavgn$. We start analysing the single average phonon vertex $\gavg$. From the definition Eq.~\eqref{eq:def_gavgn}, doing a Taylor expansion of the derivative of the Kohn-Sham potential around the equilibrium configuration $\bReq$ we obtain
\begin{align}
\gavg_{a}^{i\bk'\,m\bk}&=\Avgeq{\partial_a\Vks_{i\bk'\,m\bk}}\\
&=\partial_a\Vks_{i\bk'\,m\bk}|\pedReq
  +\frac{1}{2}\partial^3_{a\,b_1b_2}\Vks_{i\bk'\,m\bk}|\pedReq
    \Avgeq{u^{b_1}u^{b_2}}\nonumber\\
&\mkern10mu+\frac{1}{4!}\partial^5_{a\,b_1b_2b_3b_4}\Vks_{i\bk'\,m\bk}|\pedReq\Avgeq{u^{b_1}u^{b_2}u^{b_3}u^{b_4}}+\ldots\,,
\label{eq:gavg_expans_beforeWick}
\end{align}
were the sum over the repeated indices $b_h$ is understood. Since the average is done with respect to a normal distribution, we can apply Wick's theorem. Taking into account the different number of possible pairings, and from the definition Eq.~\eqref{eq:def_gn} for the standard electron-phonon vertices, we obtain
\begin{align}
\gavg_{a}^{i\bk'\,m\bk}&=g_{a}^{i\bk'\,m\bk}+3!!\,\gx{3}_{a\,b_1b_2}^{i\bk'\,m\bk}\Avgeq{u^{b_1}u^{b_2}}\nonumber\\
&\mkern10mu+5!!\,\gx{5}_{a\,b_1b_2b_3b_4}^{i\bk'\,m\bk}\Avgeq{u^{b_1}u^{b_2}}\Avgeq{u^{b_3}u^{b_4}}+\ldots\,,
\label{eq:gavg_expans}
\end{align}
where $\Avgeq{u^{b_1}u^{b_2}}=\Psi^{b_1b_2}$ is the non-interacting equal-time Green function, 
and the coefficient $(2n-1)!!$ of the $\osss{(2n-1)}{g}$ term comes from the fact that we have divided and multiplied by $(2n-1)!$, and from the $(2n-3)!!$ number of pairing combinations of $u^{a_1}\cdots u^{a_{2n-1}}$,
so that we have $(2n-1)!/(2n-2)!\times (2n-3)!!=(2n-1)!!$.  The coefficients $(2n-1)!!$ are just combinatorial prefactors that can be interpreted as the different ways of selecting one index and pairing the other $2n-2$ ones ( $(2n-1)!!=(2n-1)\times(2n-3)!!$), thus Eq.~\eqref{eq:gavg_expans} shows that the single average electron-phonon (e-ph) vertex $\gavg$ is given by the superposition of the conventional odd e-ph vertices with all the phonon legs but one closed in couples with a loop.  In Fig.~\ref{fig:gavg} the diagrammatic description of this vertex is shown, where the phonon legs paired with loops give 
a characteristic ``flower'' diagram. 

\begin{figure}[t!]
\includegraphics[width=\columnwidth]{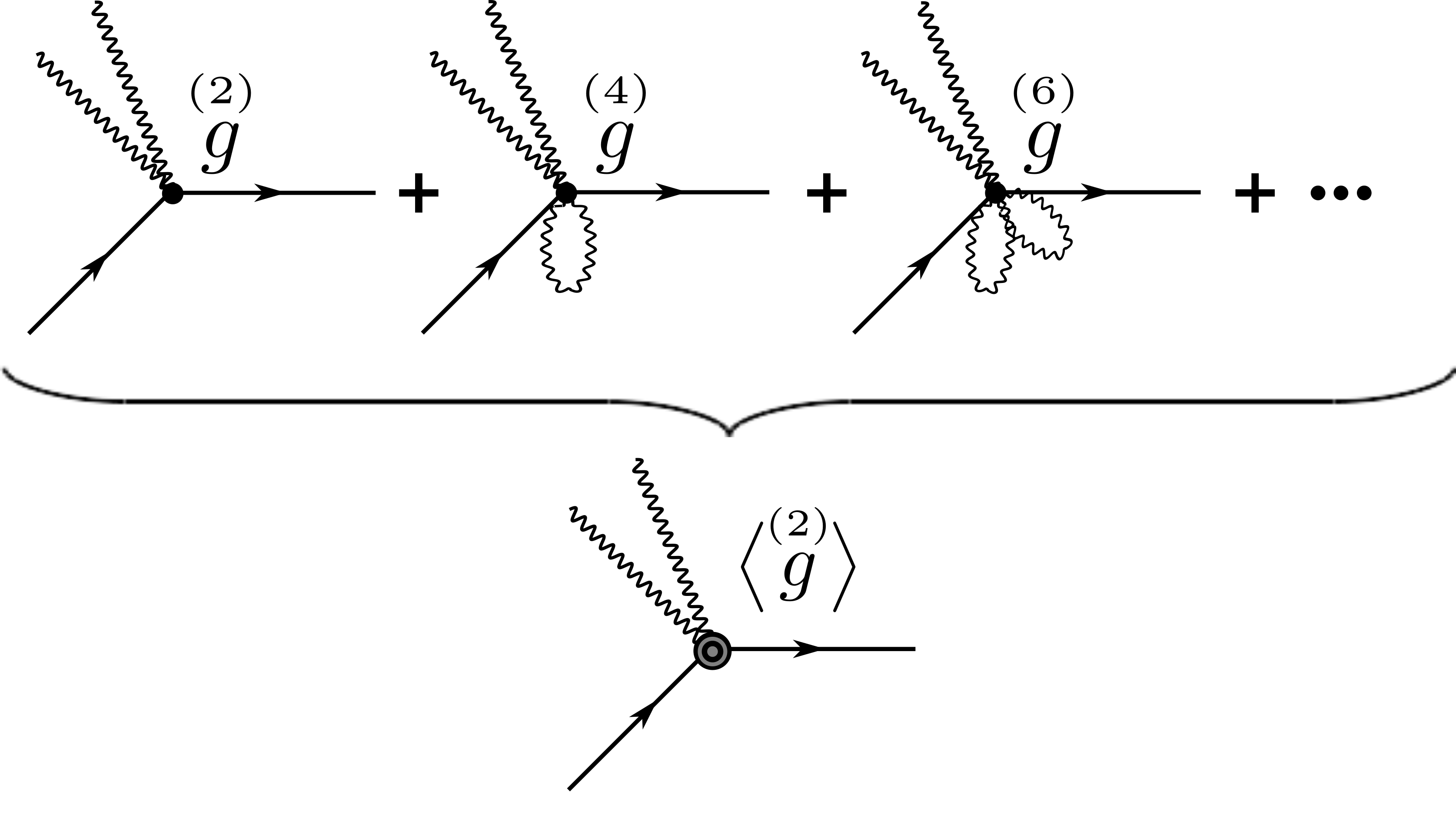}
\caption{Average double electron-phonon vertex $\gavgx{2}$, given by the superposition of double $\gx{2}$, quadrupole $\gx{4}$, sixtupole $\gx{6}$ \ldots 
phonon vertices with only one external phonon leg, and the other phonon legs paired in loops.}
\label{fig:g2avg}
\end{figure}

Along the same lines, we can analyze the average double-phonon vertex $\gavgx{2}$, first doing a Taylor expansion of the second derivative,
\begin{align}
&\gavgx{2}_{a_1a_2}^{i\bk'\,m\bk}=\frac{1}{2}\Avgeq{\partial^2_{a_1 a_2}\Vks_{i\bk'\,m\bk}}\\
&\quad=\frac{1}{2}\partial^2_{a_1 a_2}\Vks_{i\bk'\,m\bk}|\pedReq\nonumber\\
&\quad\quad+\frac{1}{2\cdot2}\partial^4_{a_1a_2\,b_1b_2}\Vks_{i\bk'\,m\bk}|\pedReq\Avgeq{u^{b_1}u^{b_2}}\nonumber\\
&\quad\quad\quad+\frac{1}{2\cdot 4!}\partial^6_{a_1a_2 b_1 b_2 b_3 b_4}\Vks_{i\bk'\,m\bk}|\pedReq\Avgeq{u^{b_1}u^{b_2}u^{b_3}u^{b_4}}+\ldots\,,
\label{eq:gavg_expans_two_beforeWick}
\end{align}
and then applying Wick's theorem,
\begin{align}
\gavgx{2}_{a_1 a_2}^{i\bk'\,m\bk}&=\gx{2}_{a_1a_2}^{i\bk'\,m\bk}+\binom{4}{2}\,\gx{4}_{a_1 a_2 b_1b_2}^{i\bk'\,m\bk}\Avgeq{u^{b_1}u^{b_2}}\nonumber\\
&\mkern10mu+\binom{6}{2}3!!\,\gx{6}_{a_1 a_2 b_1b_2b_3b_4}^{i\bk'\,m\bk}\Avgeq{u^{b_1}u^{b_2}}\Avgeq{u^{b_3}u^{b_4}}+\ldots\,,
\label{eq:gavg_expans_two}
\end{align}
where the coefficient $\binom{2n}{2}(2n-3)!!$ in front of the $\gx{2n}$ term comes from the fact that we have divided and multiplied by $(2n-2)!$, and
from the $(2n-3)!!$ ways of partitioning $u^{a_1}\cdots u^{a_{2n-2}}$ in couples. The interpretation of this formula is immediate: the average double e-ph vertex 
$\gavgx{2}$ is given by the superposition of the conventional even e-ph vertices with all the phonon legs but two paired with a loop. 
The different way of selecting the two indices of the external phonon legs is accounted by $\binom{2n}{2}$, while $(2n-3)!!$ is the number of ways the 
remaining $2n-2$ phonon legs can be paired. In Fig.~\ref{fig:g2avg} the diagrammatic description of this vertex, with the characteristic flower-type diagrams, is shown.

A similar analysis and interpretation can be done for any $n$-phonon vertex $\gavgn$, given by the superposition of the standard $(n+2h)$-phonon vertices, with $h=0,1,2,\ldots$, having all the legs, but the $n$ external ones, paired in loops. This sheds light on the definition given in Eq.~\eqref{eq:def_gavgn}: the average vertex $\gavgn$ is obtained by adding to the conventional vertex $\gn$  a correction of the second order in the atomic displacements
\begin{equation}
\gavgn^{n\bk'\,m\bk}_{a_1\cdots a_n}=\gn^{n\bk'\,m\bk}_{a_1\cdots a_n}+\Ocal(u^2)\,,
\end{equation}
which is the result of the average time-independent correction to the KS potential arising from the thermal and quantum induced fuzziness of the nuclear charge density around the equilibrium nuclear positions (this correction being given by the ``flowers'' in the diagrammatic description). 


The analysis performed in this section sheds light on the different contributions originating from the 
terms in the nuclei susceptibility formula, Eq.~\eqref{eq:chi_gauss}.  The average, flower-type, corrections to the vertices come from the Debye-Waller factors $e^{-\frac{1}{2}\bq\cdot\bPsi\cdot\bq}$ and 
$e^{-\frac{1}{2}\bq'\cdot\bPsi\cdot\bq'}$, whereas the factor $e^{\bq\cdot\bD(\tau)\cdot\bq'}$  
is responsible for the multiphonon propagator connecting two vertices. Regarding this second term, 
it is particularly convenient to work with the phonon normal coordinates $Q_{\mu}$
\begin{align}
& u^{a}=\sum_{\mu}\sqrt{\frac{1}{2M_a\omega_{\mu}}}e^a_{\mu}\,Q_{\mu}\\
& Q_{\mu}=\sum_{a}\sqrt{{2M_a\omega_{\mu}}} e^a_{\mu}\,u^a\,.
\end{align}
Indeed, the phonon Green function in normal coordinates is diagonal
\begin{align}
&\Bcal^{\mu\nu}{(i\omega_l)}=\delta^{\mu\nu}\Bcal^{\mu}\\
&\text{ with:}\nonumber\\
&\Bcal^{\mu}{(i\omega_l)}=\frac{1}{i\omega_l-\omega_\mu}-\frac{1}{i\omega_l+\omega_\mu}\,.%=\frac{2\omega_{\mu}}{(i\omega_l)^2-\omega_\mu^2}
\end{align}
Thus, we have
\begin{align}
&\Wenn_{i\bk'\, l\bk\, m \bk\, n\bk'}(i\omega_l)=\nonumber\\
&n!\sum_{\mu_1\cdots\mu_n}
\gavgn^{i\bk'\, m\bk}_{\mu_1\ldots \mu_n}
\left[(-1)^{n-1}\left(\Conv_{h=1}^{n}\Bcal^{\mu_h}\right)(i\omega_l)\right]
\gavgn^{l\bk \, n\bk'}_{\mu_1\ldots \mu_n}\,,
\label{eq:Velph_Gauss_freq_normal}
\end{align}
where
\begin{align}
\gavgn^{i\bk\, m\bk'}_{\mu_1\ldots \mu_n}&=\frac{1}{n!}\Avgeq{\frac{\partial^n\Vks_{i\bk\, m\bk'}}{\partial Q^{\mu_1}\ldots \partial Q^{\mu_n}}}\nonumber\\
&=\sum_{a_1\cdots a_n}
\frac{e^{a_1}_{\mu_1}}{\sqrt{2M_{a_1}\omega_{\mu_1}}}
\cdots 
\frac{e^{a_n}_{\mu_n}}{\sqrt{2M_{a_n}\omega_{\mu_n}}}
\,\gavgn^{i\bk\, m\bk'}_{a_1\ldots a_n}
\,.
\label{eq:def_gavgn_normal}
\end{align}
A particular useful result is obtained in the zero temperature limit $\beta\rightarrow+\infty$. 
In that case, it can be demonstrated that
\begin{equation}
\left[(-1)^{n-1}\left(\Conv_{h=1}^{n}\Bcal^{\mu_h}\right)(i\omega_l)\right]=\Bcal^{\mu_1+\cdots+\mu_n}(i\omega_l)\,,
\label{eq:Bsum_1}
\end{equation}
where
% &\Bcal^{\mu_1+\cdots+\mu_n}(i\omega_l)=\frac{2\sum_{h=1}^n\omega_{\mu_{h}}}{(i\omega_l)^2-(\sum_{h=1}^n\omega_{\mu_h})^2}
\begin{equation}
\Bcal^{\mu_1+\cdots+\mu_n}(i\omega_l)=
\frac{1}{i\omega_l-\left(\sum_{h=1}^n\omega_{\mu_h}\right)}-\frac{1}{i\omega_l+\left(\sum_{h=1}^n\omega_{\mu_h}\right)}\,,
\label{eq:Bsum_2}
\end{equation}
thus
\begin{align}
&\Wenn_{i\bk'\, l\bk\, m \bk\, n\bk'}(i\omega_l)=\nonumber\\
&\mkern80mu n!\,\sum_{\mu_1\cdots\mu_n}
\gavgn^{i\bk'\, m\bk}_{\mu_1\ldots \mu_n}
\Bigl[\Bcal^{\mu_1+\cdots+\mu_n}(i\omega_l)\Bigr]
\gavgn^{l\bk \, n\bk'}_{\mu_1\ldots \mu_n}\,.
\label{eq:Velph_Gauss_freq_normal_zeroT}
\end{align}
Later, we will employ this result in the computation of the superconducting critical temperature of conventional superconductors in the lower temperature limit.

%%%%%%%%%%%%%%%%%%%%%%%%%%%%%%%%%%%%%%%%%%%%%%%%%%%%%%%%%%%%%%%%%%%%%%%%%%%%%%%%%%%%%%%%%%%%%%%%%%%%%%%%%%%%
\section{The Gaussian $\GWen$ approximation}
\label{sec:GWen_approx}

\begin{figure}[t!]
\includegraphics[width=\columnwidth]{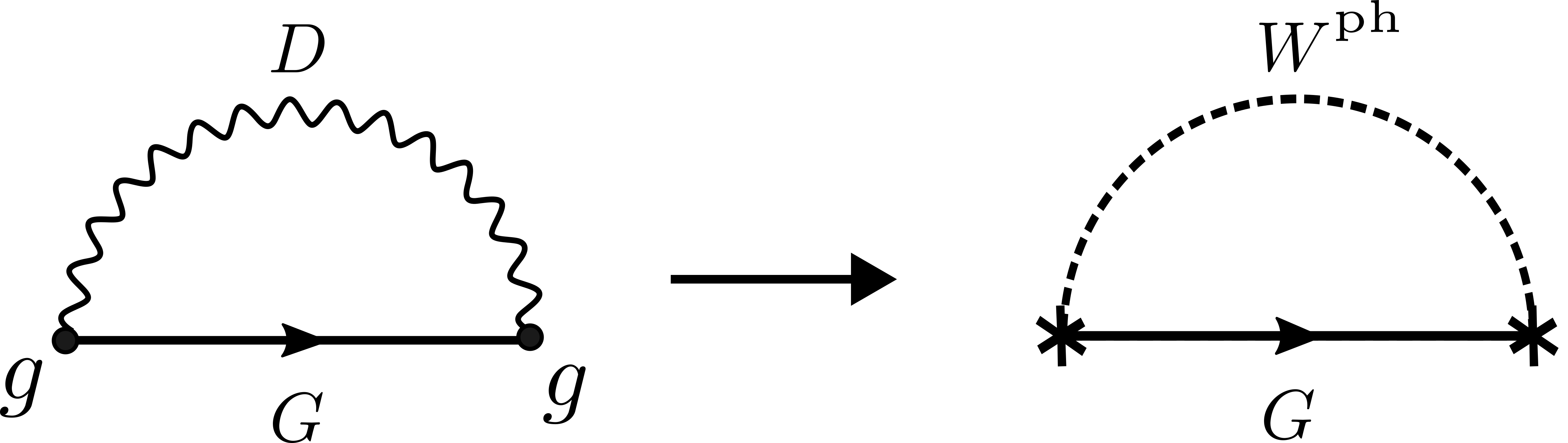}
\caption{Approach proposed to include non-linear effects in the EPI. 
The so-called Fan-Midgal (or rainbow) electron self-energy $GD$~\cite{RevModPhys.89.015003} (on the left) is replaced with the $\GWen$ self-energy (on the right). In the Gaussian approximation, the $\GWen$ self-energy  corresponds to the $GD$ self-energy plus the sum of all the  diagrams obtained connecting the two vertices with multiple phonon propagators $D$, and decorating the vertices with multiple closed Debye-Waller loops, see Fig.~\ref{fig:GWen_exp}}
\label{fig:GWen}
\end{figure}

Starting from the nuclei-mediated effective electron-electron interaction $\Wen$, we can estimate the corresponding electron self-energy by using the standard field-theory techniques. This requires the inclusion of all the possible Feynman diagrams built using the interaction $\Wen$.  However, as Midgal's ``theorem'' asserts, in the standard perturbative approach to the phonon-mediated interaction, it is appropriate to consider only the Fan-Midgal (FM) self-energy diagram, $\Sigma\simeq GD$.~\cite{migdal1958interaction,allen1983theory} Analogous reasoning behind Midgal's theorem can be qualitatively repeated in our case too, thus we consider the approximate electron self-energy $\Sigma\simeq\GWen$, see~Fig~\ref{fig:GWen}. It is worthwhile to stress that this  approximation corresponds to the nuclei-mediated-interaction counterpart of the well-known $GW$ approximation for the electron self-energy given by the screened electron-electron Coulomb interaction $W$~\cite{PhysRevB.34.5390}.

The $\GWen$ self-energy, in coordinates and Matsubara-frequency representation, is given by
\begin{equation}
\Sigma(\br,\br';ip_{h})=-\frac{1}{\beta}\sum_l\,G(\br,\br';ip_h-iw_l)\Velph(\br,\br';iw_l)\,.
\end{equation}
Using the Bloch eigenstates of the electronic problem with nuclei frozen at equilibrium, and considering, as customary, only the diagonal part of the self-energy in the band index, it is 
\begin{equation}
\Sigma_{n\bk}(ip_{h})=-\frac{1}{\beta}\!\sum_{l}\sum_{n',\bk'}G_{n'\bk'}(ip_h-iw_l)\,\Velph_{n'\bk'\,n\bk}(iw_l)\,,
\label{eq:slef_eph_bloch}
\end{equation}
where, in order to simplify the notation, 
we wrote $\Velph_{n'\bk'\,n\bk}(iw_l)$ instead of $\Velph_{n\bk\,n'\bk'\,n'\bk'\,n\bk}(iw_l)$, thus leaving the redundant identical indices understood.  Eq.~\eqref{eq:slef_eph_bloch} can be analyzed further. We consider spin-degenerate systems. Moreover, we work in a low-temperature regime, thus most of the physics happen close to the Fermi level $\ef$ (that we will set as zero of the energy). Since we are going to analyze the normal (i.e. non-superconducting) state,  following the justification given by Migdal~\cite{migdal1958interaction} and Holstein~\cite{HOLSTEIN1964410}, we consider in Eq.~\eqref{eq:slef_eph_bloch} the free-electron Green function
\begin{equation}
G_{n'\bk'}(ip_h-iw_l)=\frac{1}{ip_h-iw_l-\varepsilon_{n' \bk'}}\,.
\label{eq:free_electron_G}
\end{equation}
A crucial quantity in the study of the electron-phonon coupling is the spectral function of the nuclei-mediated electron-electron interaction, times the Fermi-energy density of states per spin, i.e. the so-called ``Eliashberg function''
(although the same name is also used in literature to indicate closely related, but different, expressions~\cite{RevModPhys.89.015003}):
\begin{equation}
\asqF_{n'\bk'\,n\bk}(\omega)=-\frac{\Nc\Nf}{\pi}\Im  \Velph_{n'\bk'\,n\bk}(\omega+i0^+)\,,
\label{eq:spectralWen}
\end{equation}
where $\Nf=\frac{1}{\Nc}\sum_{n\bk}\delta(\varepsilon_{n\bk})$ is the density of states per spin and per unit cell. Using the Eliashberg function, the ion-mediated electron-electron effective interaction is written as
\begin{align}
\Velph_{n'\bk'\,n\bk}(iw_l)&=\frac{1}{\Nc\Nf}\int_{0}^{+\infty}d\omega\,\asqF_{n'\bk'\,n\bk}(\omega)\nonumber\\
&\qquad\qquad  \left[\frac{1}{iw_l-\omega}-\frac{1}{iw_l+\omega}\right]\,.
\label{eq:Velph_w_a2F}
\end{align}
We consider the systems in the dirty/isotropic limit, meaning that we can neglect the details of $\Velph_{n'\bk'\,n\bk}(iw_l)$ w.r.t $n'\bk'$ (and $n\bk$) on constant-energy surfaces, and replace it with averages. 
Plugging Eqs.~\eqref{eq:Velph_w_a2F} and~\eqref{eq:free_electron_G} in Eq.~\eqref{eq:slef_eph_bloch}, and taking into account the isotropy when we perform the $n'\bk'$ sum,
we get in the low-temperature limit (more details in App.~\ref{sec:from_W_to_lambda})
\begin{equation}
\Sigma_{n\bk}(ip_h)=\int_0^{+\infty}d\omega\,\asqF_{n\bk}(\omega)\int_{-\infty}^{+\infty}d\varepsilon\, I[ip_h,\varepsilon,\omega]\,,
\label{eq:sigmaep_G}
\end{equation}
with
\begin{equation}
I[ip_h,\varepsilon,\omega]=\left[\frac{\Theta(\varepsilon)}{ip_h-\varepsilon-\omega}+\frac{1-\Theta(\varepsilon)}{ip_h-\varepsilon+\omega}\right]\,,
\label{eq:I}
\end{equation}
where $\Theta(\varepsilon)$ is Heaviside's step function. With $\asqF_{n\bk}(\omega)$ we are indicating the Fermi-surface (FS) average of $\asqF_{n\bk\,n'\bk'}(\omega)$ w.r.t. $n'\bk'$:
\begin{align}
\asqF_{n\bk}(\omega)
&=\vphantom{\frac{\sum_{n'\bk'}\asqF_{n'\bk'\,n\bk}(iw_l)\delta(\varepsilon_{n'\bk'})}
           {\sum_{n'\bk'}\delta(\varepsilon_{n'\bk'})}}
   \Avg{\asqF_{n' \bk'\,n\bk}}_{n'\bk'\in\text{FS}}\\
&=\frac{\sum_{n'\bk'}\asqF_{n'\bk'\,n\bk}(\omega)\delta(\varepsilon_{n'\bk'})}
{\sum_{n'\bk'}\delta(\varepsilon_{n'\bk'})}\,.
\end{align}
From the self-energy defined on the $ip_{h}$ of the imaginary axis, we can perform the analytic continuation, obtaining $\Sigma(z)\,\text{with }z\in\mathbb{C}$ and its limit to the real-frequency axis from above, $\Sigma(\Omega+i0^+)$. This allows to compute the retarded Green function and have access to plenty of information regarding electron dynamics. In particular, a quantity of significant interest in metals, even at low temperatures, is the so called ``mass-enhancement parameter'' or ``electron-phonon coupling strength'' $\lambda_{n \bk}$:
\begin{equation}
\lambda_{n \bk}=\left.-\frac{\partial \Re\Sigma_{n\bk}(\Omega+i0^+)}{\partial \Omega}\right|_{\Omega=E_{n\bk}}\,,
\label{eq:def_lambda_nk}
\end{equation}
where $E_{n\bk}$ is the energy of the quasiparticle $\Ket{n\bk}$ (at lowest order given by $\varepsilon_{n\bk}+\Re\Sigma_{n\bk}(\varepsilon_{n\bk})$). The knowledge of the $\lambda_{n \bk}$'s allows to estimate the renormalization of the bands, and in particular the correction to the electron velocity, induced by the electron-nuclei interaction. It is also related to the superconducting transition temperature of phonon-mediated superconductors. Since we are in the low-temperature limit, and we are considering
states close to the FS, we evaluate Eq.~\eqref{eq:def_lambda_nk} with $\Omega=\ef=0$, and from Eqs.~\eqref{eq:sigmaep_G} and~\eqref{eq:I} we obtain 
\begin{equation}
\lambda_{n \bk}=
\left.-\frac{\partial \Re\Sigma_{n\bk}(\Omega+i0^+)}{\partial \Omega}\right|_{\Omega=0}=
2\int_0^{+\infty}\mkern-10mu d\omega\,\frac{\asqF_{n\bk}(\omega)}{\omega}\,.
\label{eq:lambdank_T0}
\end{equation}
Since we are considering the dirty/isotropic limit, it is appropriate to consider the so-called electron-phonon ``coupling strength'' 
$\lambda$, which is the average of $\lambda_{n \bk}$ on the Fermi surface:
\begin{align}
\lambda&=\Avg{\lambda_{n \bk}}_{n\bk\in\text{FS}}\\
&=2\int_0^{+\infty}\mkern-10mu d\omega\,\frac{\asqF(\omega)}{\omega}
\end{align}
where $\asqF(\omega)=\Avg{\asqF_{n \bk}(\omega)}_{n\bk\in\text{FS}}
=\ApAvg{\asqF_{n'\bk'\,n \bk}(\omega)}_{\substack{{n'\bk'}\\{n\bk}}\in\text{FS}}$. The Eliashberg function $\asqF(\omega)$, together with the effective Coulomb potential parameter $\mu^*$~\cite{PhysRev.125.1263}, is the fundamental ingredient to compute the superconducting critical temperature $T_c$ of phonon-mediated superconductors in the isotropic limit of the Midgal-Eliashberg (ME) theory~\cite{RevModPhys.62.1027}, either by solving the ME equations or through the Allen–Dynes modified version of the McMillan semiempirical expression~\cite{PhysRevB.12.905}. 

Replacing the $GD$ expression of the self-energy with the more general $\GWen$ one, we have generalized the formulas employed in the
the isotropic approximation of the ME theory to study phonon-mediated superconductors within the Midgal approximation in the low-temperature limit (see Ref.~\citenum{allen1983theory} for a comprehensive presentation of the ME theory). Analogous generalizations can be done for the more general anisotropic case, or for finite temperature. The obtained results are general, as they are not based on any assumption about the nuclei dynamics. 

\begin{figure*}[t!]
\includegraphics[width=\textwidth]{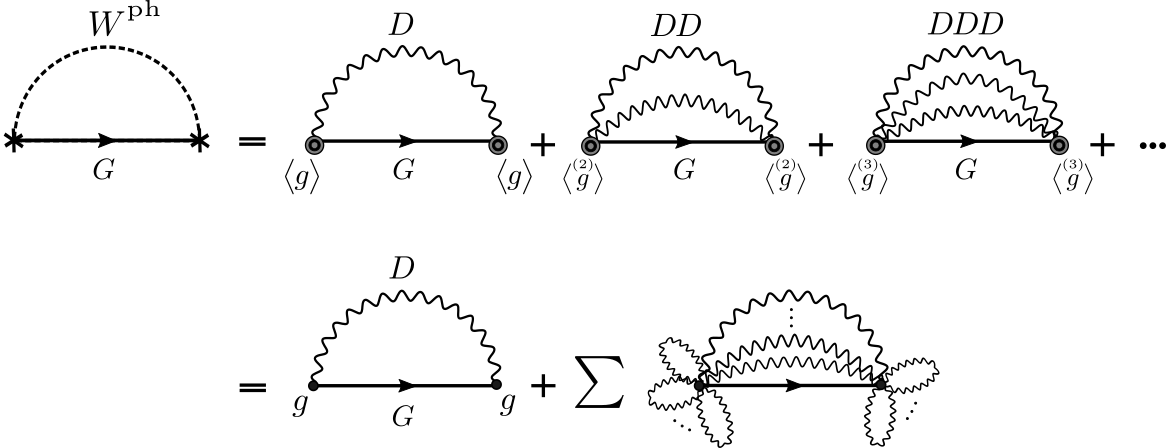}
\caption{Diagrammatic expansion of the electron self-energy in the Gaussian $G\Velph$ approximation}. 
\label{fig:GWen_exp}
\end{figure*}

In case the nuclei dynamics are described by an effective harmonic Hamiltonian, the obtained formulas can be simplified further. The crucial step is to consider in Eq.~\eqref{eq:spectralWen} the series expansion of $\Wen$ valid in the harmonic case, Eqs.~\eqref{eq:Velph_Gauss_freq_series} and~\eqref{eq:Velph_Gauss_freq_nth}, so that we can write a series expansion for $\asqF_{n'\bk'\,n\bk}(\omega)$ too:
\begin{equation}
\asqF_{n'\bk'\,n\bk}(\omega)=\sum_{n=1}^{+\infty}\overset{\sss{(n)}}{\asqF}{}_{n'\bk'\,n\bk}(\omega)\,,
\end{equation}
where the $n$-th term is given by
\begin{align}
&\overset{\sss{(n)}}{\asqF}{}_{n'\bk'\,n\bk}(\omega)=\Nc\Nf
n!\,\sum_{\mu_1\dots\mu_n}
\left|\gavgn^{n'\bk'\, n\bk}_{\mu_1\ldots \mu_n}\right|^2\nonumber\\
&\mkern100mu \times\,\frac{1}{\pi}\text{Im}\left[(-1)^{n}\left(\Conv_{h=1}^{n}\Bcal^{\mu_h}\right)(i\omega_l)\right]\,.
\end{align}
From this expansion and Eq.~\ref{eq:sigmaep_G} we obtain the corresponding expansion for the self-energy, 
whose diagrammatic representation is shown in Fig.~\ref{fig:GWen_exp}, where the relation with the Fan-Midgal $GD$ self-energy
is explicitly shown. For the zero-temperature case, from  
Eqs.~\eqref{eq:Bsum_1},~\eqref{eq:Bsum_2},~and~\eqref{eq:Velph_Gauss_freq_normal_zeroT}, in particular we have  
\begin{align}
\overset{\sss{(n)}}{\alphasqF}{}_{n'\bk'\,n\bk}(\omega)
&=\Nc\Nf\,n!  \nonumber\\
&\quad\times\sum_{\mu_1\dots\mu_n}
\left|\gavgn^{n'\bk'\, n\bk}_{\mu_1\ldots \mu_n}\right|^2
\delta\left(\omega-\sum_{h=1}^n\omega_{\mu_h}\right)\,,
\end{align}
%Of course, a series expansion follows for the other quantities, like the electron-phonon coupling strength function:
%\begin{equation}
%\lambda(\omega)=\sum_{n=1}^{+\infty}\lambdan(\omega)\,,
%\end{equation}
%with
%\begin{equation}
%\lambdan(\omega)=2\,\frac{\overset{\sss{(n)}}{\alphasqF}(\omega)}{\omega}\,,
%\end{equation}
and in the dirty/isotropic limit
\begin{align}
\overset{\sss{(n)}}{\alphasqF}(\omega)=
&\Nc\Nf\,n!\,\sum_{\mu_1\dots\mu_n}
\AAvg{\,\left|\gavgn^{n'\bk'\, n\bk}_{\mu_1\ldots \mu_n}\right|^2\,}_{\substack{{n'\bk'}\\{n\bk}}\in\text{FS}}\nonumber\\
&\mkern100mu\times\delta\left(\omega-\sum_{h=1}^n\omega_{\mu_h}\right)\,,
\label{eq:eli_notimpl}
\end{align}
i.e. the $n$th-order Eliashberg function $\overset{\sss{(n)}}{\alphasqF}(\omega)$ is proportional to the $n$-phonon joint density of states weighted by the FS-averaged $n$th-order average electron-phonon vertex. By integrating over frequency, we obtain electron-phonon coupling constant $\lambda=\sum_n\overset{{\sss{(n)}}}{\lambda}$ with
\begin{equation}
\overset{{\sss{(n)}}}{\lambda}=2\Nc\Nf n!\,\sum_{\mu_1\dots\mu_n} \frac{\AAvg{\,\left|\gavgn^{n'\bk'\, n\bk}_{\mu_1\ldots \mu_n}\right|^2\,}_{\substack{{n'\bk'}\\{n\bk}}\in\text{FS}}}{\sum_{h=1}^n\omega_{\mu_h}}.
\label{eq:lambdan_GW_gauss}
\end{equation}

In conclusion, employing the Gaussian $G\Velph$ approximation results in considering the high-order rainbow-like self-energy diagrams (i.e. the high-order FM self-energy diagrams) but with the $n$-th order Debye-Waller dressed vertices $\gavgn$ replacing the standard $n$-order bare ones $\gn$. In general, given the same phonon frequencies, this replacement is expected to suppress the  Eliashberg function and, consequently, the electron-phonon coupling~\cite{osti_6556355}. A simple way to convince ourselves about this effect is considering the fact that the average vertex is computed with $\avg{\partial^n\Vks_{n\bk\,n'\bk'}/\partial \bR^n}_{\rhoeq(\bR)}$, with $\rhoeq(\bR)$ a Gaussian distribution centered in $\bReq$ and width given by the atomic mean square displacement, and the bare vertex is obtained by replacing in this formula $\rhoeq(\bR)$ with $\delta(\bR-\bReq)$, which is $\rhoeq(\bR)$ at zero temperature in the infinite mass limit (i.e. classical limit). Therefore, the average vertex, compared to the bare one, is computed transferring weight, in the average, from the center  $\bReq$ to more distant values $\bReq+\bu$, which essentially means transferring weight from  $\left.\partial^n\Vks_{n\bk\,n'\bk'}\right|_{\bReq}$ to higher order derivatives $\left.\partial^{(n+2h)}\Vks_{n\bk\,n'\bk'}\right|_{\bReq}$ (cfr. Eqs.~\eqref{eq:gavg_expans_beforeWick} and~\eqref{eq:gavg_expans_two_beforeWick}), which are expected to be smaller. Of course, 
the larger is the atomic spread, i.e. the r.m.s of the atomic displacement (cfr. Eqs.~\eqref{eq:gavg_expans} and~\eqref{eq:gavg_expans_two}), the higher is weight transfer to higher-order derivatives, the higher is expected to be the suppression of the contribution to $\lambda$ from the average vertices.  This adds an intriguing level of complexity to the problem of the isotopic effect in the electron-phonon coupling (and thus, for example, to the related isotopic effect in superconductors' critical temperature). Looking at Eq.~\eqref{eq:lambdan_GW_gauss}, but with the vertex written in Cartesian coordinates, see Eq.~\eqref{eq:def_gavgn_normal}, we can give the estimate $\overset{{\sss{(n)}}}{\lambda}\simeq \Nf\frac{\gavgn{}^2}{M^n\overline{\omega}{}^{n+1}}$, where $\overline{\omega}$ is the average/characteristic phonon frequency, and $M$ a sort of reduced mass. We can see that now both numerator and denominator in this formula are affected by the mass, whereas in the standard approach, with $\gn$ instead of $\gavgn$, the numerator is not. For example, if we assume that the harmonic approximation for the nuclei is appropriate, we have $\overline{\omega}\simeq M^{-1/2}$, thus $\overset{{\sss{(n)}}}{\lambda}\simeq \Nf{\gavgn{}^2}/{M^{\frac{n-1}{2}}}$ and, in particular, $\overset{{\sss{(1)}}}{\lambda}\simeq \Nf{\gavg{}^2}$. If the linear-order approximation for the electron-phonon coupling is sensible, $\gavg\simeq g$, which implies that $\overset{{\sss{(1)}}}{\lambda}$ has no isotope effect, a well-know result. However, if non-linear electron-phonon coupling terms are relevant, we have a suppression of $\overset{{\sss{(1)}}}{\lambda}$ with heavier isotopes even when the nuclei dynamics is harmonic. More about isotope effect will be said in Sec.~\ref{sec:a_case_of_strongly_non-linear_electron-phonon_coupling:_palladium_hydride}.

%replacin \delta8\bu) with \bu. Replacing the delta with a more spread distribution function , is equivalnt to transfer somke weight from the n-pth order derivative to higher order ones, the sprader bwing the distribution, the more.

%%%%%%%%%%%%%%%%%%%%%%%%%%%%%%%%%%%%%%%%%%%%%%%%%%%%%%%%%%%%%%%%%%%%%%%%%%%%%
\section{The first-principles implementation}
\label{sec:the_first-principles_implementation}

In this section we describe the computational implementation of our new theory used to perform \textit{ab initio} calculations.
We considered the effective harmonic case, thus we assume that we have access to a harmonic dynamic theory to describe nuclei dynamics, resulting from the standard harmonic approximation or from an effective harmonic theory that includes anharmonic effects like  the SCHA. It is worthwhile to stress that, in principle, even when the harmonic approximation applies well, non-linear electron-phonon coupling effects can be non-negligible and may need to be included, cfr. end of Sec.~\ref{sec:the_linear_order}. 
The implemented method is based on a real-space stochastic approach. By diagonalizing the real-space effective harmonic dynamical matrix $D_{ab}=\phi_{ab}/\sqrt{M_aM_b}$, we obtain phonon frequencies $\omega_{\mu}$ and eigenmodes $e^{\mu}_a$, and the equilibrium nuclei probability distribution function $\rhoeq(\bu)$, Eq.~\eqref{eq:rhoeqharm}.
The key quantities that we need to compute are the average vertices $\gavgn$, Eq.~\eqref{eq:def_gavgn}. To do that we use a stochastic approach
using an ensemble of $\Ncal$ distorted atomic configurations $\bu_{\sss{(\Ical)}}$ generated according to the distribution $\rhoeq(\bu)$.
Of course, a straightforward Monte Carlo-like evaluation of the average e-ph vertices would be impracticable, since it would  require to compute the $n$-th order derivative of the KS potential for each element $\bu_{\sss{(\Ical)}}$ of the ensemble, which is an extraordinary time-consuming task. However, since we are considering a normal distribution, the evaluation of Eq.~\eqref{eq:def_gavgn} can be done through the sole knowledge of the distorted potentials, not its derivatives, and their matrix elements with the undistorted eigenfunctions $\Vks_{i\bk'\,m\bk}(\bu_{\sss{(\Ical)}})=\Braket{i\bk'|\Vkshat(\bu_{\sss{(\Ical)}})|m\bk}$ (i.e. only the Kohn-Sham potential depends on the distortion $\bu$, not the eigenstates $\Ket{m\bk}$).

% the matrix-element potentials $\Vks_{i\bk'\,m\bk}(\bu_{\sss{(\Ical)}})=\Braket{i\bk'|\Vkshat(\bu_{\sss{(\Ical)}})|m\bk}$, i.e. we need to compute only the distorted potentials, not its derivatives (we remark, once again, that the KS matrix elements are done w.r.t the eigenfunctions $\Ket{m\bk}$ of the undisorted atmonic configuration, . 

Indeed, given a generic function $\Ocal(\bu)$, from Eqs.~\eqref{eq:rhoeqharm} and~\eqref{eq:Psi_def}, using integration by parts we can write
\begin{equation}
\Avgeq{\partial_a\Ocal(\bu)}=\Avgeq{U_a\Ocal(\bu)}\,,
\end{equation}
where 
\begin{align}
&U_a=\sum_b\Psi^{-1}_{ab}u^b\qquad\text{ with:}\\
&\Psi^{-1}_{ab}=\sum_{\mu}\frac{2\omega_{\mu}}{1+2n_\mu}\sqrt{M_a}e^{a}_{\mu}\sqrt{M_b}e^{b}_{\mu}\,,
\end{align}
so that, in particular, we have 
\begin{equation}
\gavg_a^{i\bk'\,m\bk}=\Avgeq{\partial_a \Vks_{i\bk'\,m\bk}}=\Avgeq{U_a\Vks_{i\bk'\,m\bk}}\,.
\label{eq:gavg_sto}
\end{equation}
Applying the same formula iteratively, and using the fact that $\partial_aU_b=\Psi^{-1}_{ab}$,
we obtain the higher-order derivatives like
\begin{align}
\gavgx{2}_{a b}^{i\bk'\,m\bk}&=\Avgeq{\partial^2_{a b}\,\Vks_{i\bk'\,m\bk}}\nonumber\\
&=\Avgeq{U_aU_b\Vks_{i\bk'\,m\bk}}\nonumber\\
&\mkern20mu +
\contraction{\,}{U_a}{}{U_b}
\Avgeq{ U_a U_b\,\Vks_{i\bk'\,m\bk}}\,,
\label{eq:gavg2_sto}
\end{align}
where $\contraction{\!}{U_a}{}{U_b} U_a U_b=-\Psi^{-1}_{ab}$, 
\begin{align}
\gavgx{3}_{a b c}^{i\bk'\,m\bk}&=\Avgeq{\partial^3_{a b c}\,\Vks_{i\bk'\,m\bk}}\nonumber\\
&=\Avgeq{U_aU_bU_c\Vks_{i\bk'\,m\bk}}\nonumber\\
&\mkern20mu +\sum_{\sqcup}
\contraction{\,}{\,U_a}{U_b}{U_c}
\Avgeq{ U_a U_bU_c\,\Vks_{i\bk'\,m\bk}}\,,
\label{eq:gavg3_sto}
\end{align}
where $\displaystyle{\sum_{\sqcup}}$ is the sum over all the possible contractions~$\sqcup$,
\begin{align}
\gavgx{4}_{a b c d}^{i\bk'\,m\bk}&=\Avgeq{\partial^4_{a b c d}\,\Vks_{i\bk'\,m\bk}}\nonumber\\
&=\Avgeq{U_aU_bU_cU_d\Vks_{i\bk'\,m\bk}}\nonumber\\
&\mkern20mu +\sum_{\sqcup}
\contraction{\,}{\,U_a}{U_b}{U_c}
\Avgeq{ U_a U_bU_cU_d\,\Vks_{i\bk'\,m\bk}}
\nonumber\\
&\mkern40mu +\sum_{\sqcup\sqcup}
\contraction{\,}{\,U_a}{U_b}{U_c}
\contraction[2ex]{\,U_a}{\,U_b}{U_c}{U_d}
\Avgeq{ U_a U_bU_cU_d\,\Vks_{i\bk'\,m\bk}}\,,
\label{eq:gavg4_sto}
\end{align}
and $\displaystyle{\sum_{\sqcup \sqcup}}$ is the sum over all the possible double contractions,
\begin{align}
\gavgx{5}_{a b c d e}^{i\bk'\,m\bk}&=\Avgeq{\partial^5_{a b c d e}\,\Vks_{i\bk'\,m\bk}}\nonumber\\
&=\Avgeq{U_aU_bU_cU_dU_e\Vks_{i\bk'\,m\bk}}\nonumber\\
&\mkern20mu +\sum_{\sqcup}
\contraction{\,}{\,U_a}{U_b}{U_c}
\Avgeq{ U_a U_bU_cU_dU_e\,\Vks_{i\bk'\,m\bk}}
\nonumber\\
&\mkern40mu +\sum_{\sqcup\sqcup}
\contraction{\,}{\,U_a}{U_b}{U_c}
\contraction[2ex]{\,U_a}{\,U_b}{U_c}{U_d}
\Avgeq{ U_a U_bU_cU_dU_e\,\Vks_{i\bk'\,m\bk}}\,,
\label{eq:gavg5_sto}
\end{align}
and so forth. Therefore, in order to compute the average vertices $\gavgn$ using equations such as Eqs.~\eqref{eq:gavg_sto}~--~\eqref{eq:gavg5_sto}, given an ensemble of $\Ncal$ distorted atomic configurations  $\bu_{\sss{(\Ical)}}$ generated according the distribution $\rhoeq(\bu)$, we just need to compute $\Vkshat(\bu_{\sss{(\Ical)}})$ and the matrix elements $\Vks_{i\bk'\,m\bk}(\bu_{\sss{(\Ical)}})$ with the undistorted wave functions, and estimate the averages
\begin{align}
&\Avgeq{\Vks_{i\bk'\,m\bk}}\simeq\frac{1}{\Ncal}\sum_{\Ical=1}^{\Ncal}\Vks_{i\bk'\,m\bk}(\bu_{\sss{(\Ical)}})\,,\\
&\Avgeq{U_a\Vks_{i\bk'\,m\bk}}\simeq\frac{1}{\Ncal}\sum_{\Ical=1}^{\Ncal}U_a^{\!\sss{(\Ical)}}\,\Vks_{i\bk'\,m\bk}(\bu_{\sss{(\Ical)}})\,,\\
&\Avgeq{U_aU_b\Vks_{i\bk'\,m\bk}}\simeq\frac{1}{\Ncal}\sum_{\Ical=1}^{\Ncal}U_a^{\!\sss{(\Ical)}}U_b^{\!\sss{(\Ical)}}\,\Vks_{i\bk'\,m\bk}(\bu_{\sss{(\Ical)}})\,,\\
&\mkern80mu\ldots\nonumber
\end{align}
which are obviously exact in the $\Ncal\rightarrow+\infty$ limit.

As explained in the previous sections, in the adopted formalism the undistorted crystal is described in terms of a macroscopic PBCs supercell made of $N_c$ unit cells, and this constitutes the basis for the description of the electronic Bloch eigenstates $\Ket{n\bk}$ of the undistorted crystal. In principle, the considered distorted atomic configurations $\bR=\bReq+\bu$, distributed according to $\rhoeq(\bu)$, are totally general (i.e. they are  periodic only over the macroscopic PBCs supercell). However, as it is customary in standard harmonic phonon calculations too, we consider distorted atomic configurations periodic on a supercell large enough to reach the convergence (thermodynamic limit), but definitely smaller than the macroscopic PBCs supercell. 
Since we are considering a real-space approach, we consider this supercell as the ``unit cell'' (uc) of the crystal, and define the electronic Bloch eigenstates of the undistorted atomic configuration accordingly. In other words, we describe the equilibrium system in a (large) supercell - which is the ``unit cell'' in our description in terms of the Bloch pseudomomentum, and we consider only atomic distortions periodic on the unit cell (i.e. we consider $\bq=0$, i.e. $\Gamma$, atomic distortions only). Therefore, we consider atomic configurations
$\bR$ such that
\begin{align}
\Vks_{i\bk'\, m\bk}(\bR)&=\Braket{i\bk'|\Vkshat(\bR)|m\bk }\\
&=\delta_{\bk\bk'}\Braket{i\bk|\Vkshat(\bR)|m\bk }\\
&=\delta_{\bk\bk'}\Vks_{i m\bk}(\bR)\,,
\end{align}
where in the last line we have used a simplified notation to get rid of the redundant double $\bk$ index (similarly, in all the quantities with two identical $\bk$ indices, only one of them will be displayed), and
\begin{align}
\Vks_{i m\bk}(\bR)
&=\int d\br \,\psi_{i\bk}^*(\br)\,\Vks(\br,\bR)\, \psi_{m\bk}(\br)\\
&=\int_{\text{uc}}d\br \,u_{i\bk}^*(\br)\,\Vks(\br,\bR)\, u_{m\bk}(\br) .
\end{align}

In summary, in order to perform the calculations we start with a supercell of the system (large enough to obtain converged results) and, as a first thing,
for the undistorted atomic configuration, using a $\bk$-grid fine enough to reach converged results, we compute the electronic eigenfunctions $\Ket{n\bk}$ and energies $\epsilon_{n\bk}$, the Fermi energy $\ef$, and the Fermi-level density of states per cell $\Nf=\frac{1}{\Nc}\sum_{n\bk}\delta(\varepsilon_{n\bk})$. 
After that, using the dynamical matrix $D_{ab}=\phi_{ab}/\sqrt{M_aM_b}$ defined in the supercell, we compute frequencies $\omega_{\mu}$ and eigenmodes $e_{\mu}^a$, and generate an ensemble of $\NI$ distorted atomic configurations $\bRI=\bReq+\buI$ distributed according to the corresponding normal distribution $\rhoeq(\bu)$. For each element of the ensemble, we compute the KS potential $\Vkshat(\buI)$ and the braket with the equilibrium electronic wave-functions, $\Vks_{nm\,\bk}(\buI)=\Braket{i\bk|\Vkshat(\buI)|m\bk }$.
% 
% We also assume that we have access to a harmonic dynamic theory to describe nuclei dynamics, By diagonalizing the real-space harmonic dynamical matrix $D_{ab}=\phi_{ab}/\sqrt{M_aM_b}$, we obtain phonon frequencies $\omega_{\mu}$ and eigenmodes $e^{\mu}_a$, and the equilibirium nuclei probability distribution function $\rhoeq(\bu)$ (since we are in a supercell, i.e. we are considering $\Gamma$ distortions, we don't have phonon-momentum $\bq$ dependence in the dynamical matrix and in the corresponding frequencies and eigenmodes). Using $\rhoeq(\bu)$, we can generate a certain number $\NI$ of configurations  $\bRI=\bReq+\buI$ and, for each of the, compute the KS potential $\Vkshat(\buI)$ and the braket with the equilibriu electronic wave-functions $\Ket{n\bk}$, $\Vks_{nm\,\bk}(\buI)=\Braket{i\bk|\Vkshat(\buI)|m\bk }$. 
Using these quantities, we compute the averages $\Avgeq{\Vks_{nm\,\bk}}$, $\Avgeq{U^a\Vks_{nm\,\bk}}$, $\Avgeq{U^aU^b\Vks_{nm\,\bk}}$ \textellipsis, and from them the average vertices $\gavg_a^{i m\,\bk}$, $\gavgx{2}_{ab}^{i m\,\bk}$, $\gavgx{3}_{abc}^{i m\,\bk}$, etc. Once the average vertices are computed, we have all the ingredients to compute the Eliashberg function 
\begin{equation}
{\alphasqF}(\omega)=\sum_n\overset{\sss{(n)}}{\alphasqF}(\omega)\,,
\end{equation}
where
\begin{align}
&\overset{\sss{(n)}}{\alphasqF}(\omega)=
n! \sum_{\mu_1\dots\mu_n} \delta\left(\omega-\sum_{h=1}^n\omega_{\mu_h}\right)\nonumber\\
&\times\left[\frac{1}{\Nf}\frac{1}{\Nc}\sum_{\bk\,nm}{}^{\!\!'}\delta\left(\epsilon_{m\bk}-\ef\right)
\delta\left(\epsilon_{n\bk}-\ef\right)   
\left|\gavgn^{mn\bk}_{\mu_1\ldots \mu_n}\right|^2\right] 
\,,
\label{eq:eli_impl}
\end{align}
and the quantities obtainable from it, like the electron-phonon coupling constant
\begin{equation}
\lambda=\sum_n\,\overset{\sss{(n)}}{\lambda} ,
\end{equation}
where
\begin{align}
\overset{\sss{(n)}}{\lambda}=
&\frac{2 n!}{\Nf \Nc} \sum_{\mu_1\dots\mu_n}\left(\sum_{h=1}^n\omega_{\mu_h}\right)^{-1}\nonumber\\
&\quad\times
\left[\sum_{\bk\,nm}{}^{\!\!'}\delta\left(\epsilon_{m\bk}-\ef\right)
\delta\left(\epsilon_{n\bk}-\ef\right)   
\left|\gavgn^{mn\bk}_{\mu_1\ldots \mu_n}\right|^2\right] 
\label{eq:lambdan_compt}
\end{align}

Notice that Eq.~\eqref{eq:eli_impl} has been obtained from Eq.~\eqref{eq:eli_notimpl} by explicitly writing the double FS average, taking into account
that $\gavgn^{n'\bk'\, n\bk}_{\mu_1\ldots \mu_n}=\delta_{\bk\bk'}\gavgn^{n'\!n\,\bk}_{\mu_1\ldots \mu_n}$, and dividing the numerator and the denominator
by $\Nc$. Moreover, we have indicated with 
with $\sum_{\bk\,n,m}'$ the restricted sum
\begin{equation}
\sum_{\bk,n,m}{}^{\!\!'}=\sum_{\bk\neq 0}\sum_{\substack{n, m\\ \,\enex{\bk n}\neq\enex{\bk m}}}
\end{equation}
which we employ in order to exclude terms which are negligible in the converged result, but which can be a cause of numerical instabilities in approximated calculations and dramatically hampers the convergence with respect to the supercell size (more details in App.~\ref{sec:restricted_sum}). 
%%%%%%%%%%%%%%%%%%%%%%%%%%%%%%%%%%%%%%%%%%%%%%%%%%%%%%%%%%%%%%%%%%%%%%%%%%%%%

%%%%%%%%%%%%%%%%%%%%%%%%%%%%%%%%%%%%%%%%%%%%%%%%%%%%%%%%%%%%%%%%%%%%%%%%%%%%%
\section{A case of strongly non-linear electron-phonon coupling: palladium hydride}
\label{sec:a_case_of_strongly_non-linear_electron-phonon_coupling:_palladium_hydride}

\begin{figure*}[t!]
\includegraphics[width=\textwidth]{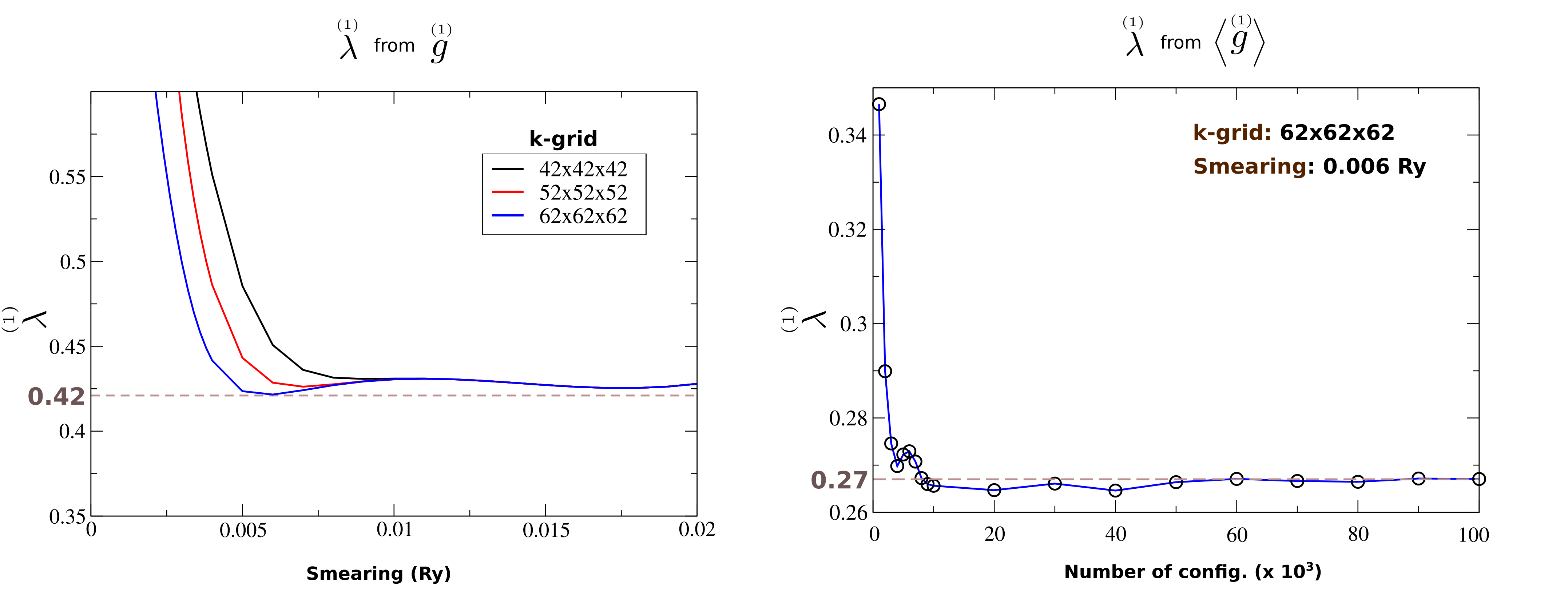}
\caption{Convergence study of $\overset{\sss{(1)}}{\lambda}$ for PdH computed by employing Eq.~\eqref{eq:lambdan_compt}.
Left-hand panel: value of $\overset{\sss{(1)}}{\lambda}$ with $\overset{\sss{(1)}}{g}$ computed by finite differences, shown as a function of the k-point grid used to perform the Brillouin-zone sum and of the Gaussian smearing employed to model the delta functions. The converged value is estimated using a $62\times 62\times 62$ grid and a smearing of $0.006\,\text{Ry}$.
Right-hand panel: $\overset{\sss{(1)}}{\lambda}$ computed using $\langle\overset{\sss{(1)}}{g}\rangle$, evaluated with the same k-point grid and smearing parameters. The results are shown as a function of the population size used to stochastically compute the average $\langle\overset{\sss{(1)}}{g}\rangle$, following the approach described in Section~\ref{sec:the_first-principles_implementation}.}. 
\label{fig:Lambda1_FD}
\end{figure*}
\begin{figure*}[t!]
\includegraphics[width=\textwidth]{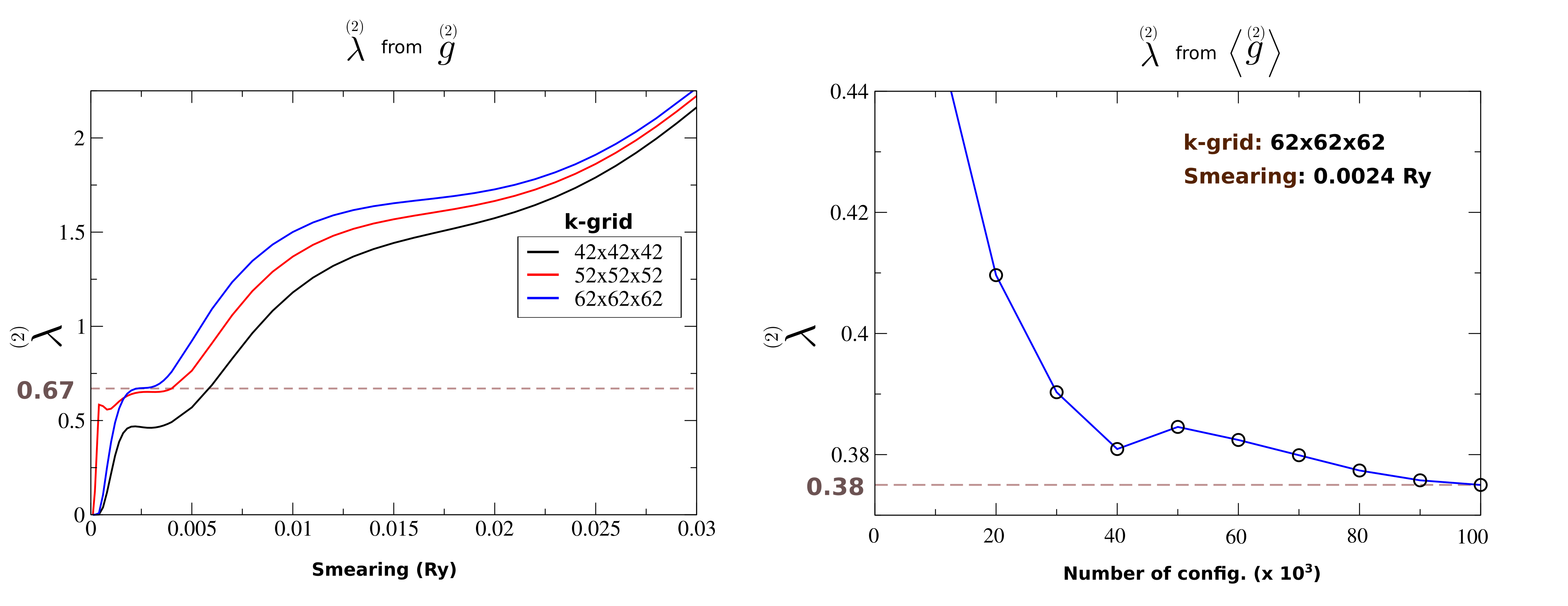}
\caption{Convergence study of $\overset{\sss{(2)}}{\lambda}$ for PdH computed by employing Eq.~\eqref{eq:lambdan_compt}.
Left-hand panel: value of $\overset{\sss{(2)}}{\lambda}$ with $\overset{\sss{(2)}}{g}$ computed by finite differences, shown as a function of the k-point grid used to perform the Brillouin-zone sum and of the Gaussian smearing employed to model the delta functions. The converged value is estimated using a $62\times 62\times 62$ grid and a smearing of $0.00024\,\text{Ry}$.
Right-hand panel: $\overset{\sss{(2)}}{\lambda}$ computed using $\langle\overset{\sss{(2)}}{g}\rangle$, evaluated with the same k-point grid and smearing parameters. The results are shown as a function of the population size used to stochastically compute the average $\langle\overset{\sss{(2)}}{g}\rangle$, following the approach described in Section~\ref{sec:the_first-principles_implementation}.}. 
\label{fig:Lambda2_FD}
\end{figure*}

Palladium hydride, PdH, is a conventional phonon-mediated superconductor, which displays the largest isotope effect anomaly known in literature: the heavier the isotope, the larger the  superconducting critical temperature~\cite{Stritzker1974,PhysRevB.10.3818,SCHIRBER1984837}. There is a broad consensus that the strong anharmonicity of PdH is at the origin of this anomaly. The reason is that, as also shown in Sec.~\ref{sec:GWen_approx}, at linear order $\lambda\propto \Nf\,g^2/M\langle \omega^2\rangle$, where $\Nf$ is the Fermi-level density of states, $g$ is the linear electron-phonon coupling vertex (mass independent), $\overline{\omega}$ is the average/characteristic phonon frequency, and $M$ is the reduced atomic mass. At harmonic level the frequency goes like $M^{-\frac{1}{2}}$, thus $\lambda$ is not affected by the change of mass, thus the superconducting critical temperature $T_c\simeq\overline{\omega}\, e^{-1/\lambda}$ is smaller for the heavier isotope~\cite{PhysRev.167.331}. However, in palladium hydride anharmonicity renormalizes the phonon frequencies to higher values, more for PdH than for PdD, and this difference compensates and exceeds the difference of the mass $M$ between H and D, making $\lambda$ so much smaller in PdH than in PdD to make $T_c\simeq\overline{\omega}\, e^{-1/\lambda}$ larger for the heavier isotope~\cite{Ganguly1973,allen1983theory,ErreaB2013}. For this reason, PdH is the ideal candidate to verify what happens when non-linear e-ph effects are included in the analysis of systems with strongly anharmonic nuclei dynamics. 

We performed density-functional theory (DFT) calculations using the QUANTUM-ESPRESSO package~\cite{Giannozzi2009}, and we employed optimized norm-conserving pseudopotentials (ONCCPSP)~\cite{PhysRevB.88.085117,VANSETTEN201839}, together with the Perdew–Burke–Ernzerhof (PBE) parametrization of the generalized gradient approximation (GGA) for the exchange–correlation functional~\cite{PhysRevLett.77.3865}. Notice that the use of ultrasoft pseudopotentials would require a modification of the equations presented here to account for the presence of augmentation charges. Calculations were performed on a $2\times 2 \times 2$ supercell. A plane-wave cutoff of $65$~Ry was used, together with a $12 \times 12 \times 12$ k-point mesh for the Brillouin-zone integrals required to compute the Kohn–Sham potentials. Due to the strongly anharmonic character of PdH (the system is not even stable at the harmonic level), we employed the SSCHA to obtain an effective harmonic description of the nuclear dynamics~\cite{ErreaB2013}. First, we computed $\lambdax{1}$ and $\lambdax{2}$ through the $GD$ and $GDD$ self-energies, using “bare”, i.e. non-averaged, phonon vertices $g_{a}^{i\bk',m\bk}$ and $\gx{2}_{a b}^{i\bk',m\bk}$, which were obtained by evaluating the first and second derivatives of the Kohn–Sham potential via finite differences.

The choice of the k-point grid used to perform the Brillouin-zone sum and of the Gaussian smearing employed to model the delta functions in Eq.~\eqref{eq:lambdan_compt} for the calculation of $\lambdax{1}$ and $\lambdax{2}$ was guided by a convergence analysis, whose results are shown in the left-hand panels of Figs.~\ref{fig:Lambda1_FD}~and~\ref{fig:Lambda2_FD}. The obtained results, reported in Tab.~\ref{tab:PdH}, confirm the expectations: the second-order contribution $\lambdax{2}$ is not negligible when compared to the first-order term $\lambdax{1}$. This indicates that, in the absence of other compensating effects, restricting the electron–phonon coupling calculations for PdH to the linear order is not an appropriate approach. This may be at the origin of the poor agreement between the superconducting critical temperature $T_c$ computed within the linear approximation and the experimental value, even when anharmonic corrections are included in the phonon propagator (with an error of the order of $\sim 40$)~\cite{ErreaB2013}.

We then computed $\lambdax{1}$ and $\lambdax{2}$ using the averaged vertices $\gavg_{a}^{i\bk'\,m\bk}$ and $\gavgx{2}_{a b}^{i\bk'\,m\bk}$, respectively, which were obtained following the stochastic approach described in Section~\ref{sec:the_first-principles_implementation}, and employing the same k-point grid and smearing parameter used for the calculation of the corresponding quantities with the bare vertices. The convergence of the results with respect to the size of the sampling population used to compute the averaged vertices is shown in the right-hand panels of Figs.~\ref{fig:Lambda1_FD}~and~\ref{fig:Lambda2_FD}. As expected (see Sec.~\ref{sec:GWen_approx}), the Debye–Waller vertex renormalization leads to a significant suppression of the electron–phonon coupling, as reported in Tab.~\ref{tab:PdH}.

From the analysis presented above, it emerges that an accurate treatment of electron–phonon coupling effects in systems with nonlinear EPI relies on two competing ingredients contributing with opposite signs. The inclusion of higher-order FM-like diagrams leads to positive corrections to the linear-order coupling, whereas the use of averaged vertices instead of bare ones suppresses these contributions. A more accurate analysis of the non-linear electron-phonon coupling effects in PdH and PdD is presented in Ref.~, where it is also shown that to the inverse isotope effect in this system contributes not only the phonon-frequency anharmonic renormalization, but also the mass-dependent suppression of the electron-phonon coupling caused by the average vertices (higher with deuterium than with hydrogen). 
\begin{table}[t!]
\begin{center}
\begin{ruledtabular}
\begin{tabular}{ccc}
\rule{0pt}{2ex}    
 & \text{with }$\gn$ & \text{with }$\gavgn$ \rule[-1.6ex]{0pt}{0ex}    \\
\hline\\
$\overset{\sss{(1)}}{\lambda}$ &  0.42  &  0.27 \\
$\overset{\sss{(2)}}{\lambda}$ &  0.67  &  0.38 
\end{tabular}
\end{ruledtabular}
\end{center}
\caption{Comparison between $\overset{\sss{(1)}}{\lambda}$ and $\overset{\sss{(2)}}{\lambda}$ calculated for PdH with the $GD$ and the $GDD$ self-energy, respectively, employing the non-averaged vertices $\gn$ and the averaged vertices $\gavgn$.}
\label{tab:PdH}
\end{table}
%%%%%%%%%%%%%%%%%%%%%%%%%%%%%%%%%%%%%%%%%%%%%%%%%%%%%%%%%%%%%%%%%%%%%%%%%%%%%%%%%%%%%%%%%%%%%%%%%%%%%%%%%%%%%%%%%%%%%%%%%%%%%%%%%%%

%%%%%%%%%%%%%%%%%%%%%%%%%%%%%%%%%%%%%%%%%%%%%%%%%%%%%%%%%%%%%%%%%%%%%%%%%%%%%
\section{A case of linear electron-phonon coupling: aluminum}
\label{sec:a_case_of_linear_electron-phonon_coupling:_aluminum}
Having discussed a system characterized by strong non-linear electron–phonon effects, we now turn to a complementary case in which such effects are expected to be negligible, so as to assess whether the proposed method correctly recovers the standard linear result. Aluminum is ideally suited for this purpose. Indeed, Al is a superconductor with a critical temperature of 1.18~K at zero pressure~\cite{PhysRevLett.35.104} and is well known to be accurately described within the harmonic approximation for the nuclear dynamics and the linear approximation for the electron–phonon coupling~\cite{PhysRevLett.96.047003}. In this case, as discussed in Sec.~\ref{sec:the_linear_order}, one therefore expects that the averaged electron–phonon matrix elements essentially coincide with the standard ones, $\gavgn_{a_1\cdots a_n}^{i\bk'\,m\bk}\simeq \gn_{a_1\cdots a_n}^{i\bk'\,m\bk}$, and that the Fan–Migdal $GD$ self-energy provides a good approximation to the full $G\Wen$ self-energy (so that, in particular, $\lambda\simeq\overset{\sss{(1)}}{\lambda}$).

To explicitly verify that the method satisfies this requirement, we performed simulations applying the same computational strategy adopted for PdH to the case of Al. Calculations were carried out using the Perdew–Zunger local-density approximation~\cite{PhysRevB.23.5048}. Since this system has already been extensively studied in the literature, with results in excellent agreement with experiments, and since the present calculations are meant solely as a benchmark rather than for quantitative comparison with experimental data, in order to limit the computational cost we did not perform a systematic convergence analysis with respect to the choice of pseudopotentials, plane-wave cutoff, $\bk$-point mesh for the Brillouin-zone integration, or the supercell size employed in the stochastic evaluation of averages, ~\footnote{Calculations have been performed on a $2\times 2 \times 2$ supercell. A $20$~Ry cutoff is used for the plane-wave basis and a $4 \times 4 \times 4$ mesh for the Brillouin-zone integrals to compute the Kohn–Sham potentials. For the average on the Fermi surface, we computed the undistorted wave functions with $\bk$ points on a $60 \times 60 \times 60$ grid.}.

Within this setup, we computed $\overset{\sss{(1)}}{\lambda}$ and $\overset{\sss{(2)}}{\lambda}$ using the $GD$ and $GDD$ self-energies, respectively, employing both bare and averaged vertices. As reported in Table~\ref{tab:aluminum}, the results fully confirm the expectations for a system in the linear regime: the values obtained using averaged or bare vertices are essentially identical, and the second-order contribution is negligible with respect to the first-order one.

\begin{table}[t!]
\begin{center}
\begin{ruledtabular}
\begin{tabular}{ccc}
\rule{0pt}{2ex}    
 & \text{with }$\gn$ & \text{with }$\gavgn$ \rule[-1.6ex]{0pt}{0ex}    \\
\hline\\
$\overset{\sss{(1)}}{\lambda}$ & 0.297  & 0.287 \\
$\overset{\sss{(2)}}{\lambda}$ & 0.007  & 0.007 
\end{tabular}
\end{ruledtabular}
\end{center}
\caption{Comparison between $\overset{\sss{(1)}}{\lambda}$ and $\overset{\sss{(2)}}{\lambda}$ calculated for Al with the $GD$ and the $GDD$ self-energy, respectively, employing the non-averaged vertices $\gn$ and the averaged vertices $\gavgn$.}
\label{tab:aluminum}
\end{table}
%%%%%%%%%%%%%%%%%%%%%%%%%%%%%%%%%%%%%%%%%%%%%%%%%%%%%%%%%%%%%%%%%%%%%%%%%%%%%

%%%%%%%%%%%%%%%%%%%%%%%%%%%%%%%%%%%%%%%%%%%%%%%%%%%%%%%%%%%%%%%%%%%%%%%%%%%%%%%%%%%%%%%%%%%%%%%%%%%%%
\section{Conclusions}
\label{sec:conclusions} 

In this work we present a non-perturbative method to include non-linear effects in the calculation of the EPI from first principles. The method is based on the definition of the nuclei-mediated effective electron-electron interaction  $\Wen$. The electrons are treated at mean-field Kohn-Sham level, and the nuclei dynamics are described at an effective harmonic level, which can include anharmonic effects  within the self-consistent harmonic approximation. The electron self-energy and related quantities are computed within the $\GWen$ approximation, the first term in the expansion of the self-energy operator in terms of the effective electron-electron interaction $\Wen$ and the electron Green’s function $G$. The pivotal quantities used in this method are the thermal/quantum averages of the EPI vertices, which are computed within a real-space stochastic approach in supercell. The \emph{ab initio} implementation of the method only requires the calculation of the Khon-Sham potential of random configurations created according to the nuclear probability distribution function. In order to validate the method, and demonstrate its capabilities, we employed the $G\Wen$ method with two superconductors, aluminum and palladium hydride, and compared the obtained results with the results of standard linear EPI calculations. Aluminum is a harmonic system, with weak EPI and for which non-linear EPI terms are considered negligible. The calculations confirm this, showing that the results obtained applying the new method coincide with the ones obtained with the standard linear EPI technique. On the contrary, palladium hydride is strongly anharmonic. Our calculations show that non-linear EPI effects are relevant in this system, and that the usage of the $\GWen$ approximation significantly corrects the results obtained within the linear EPI approach. 

The presented method joins other methods recently developed to take into account EPI effects beyond the lowest perturbative order. Compared to other method exclusively based on statistical averaging~\cite{PhysRevB.102.045126,PhysRevB.97.115145}, our $\GWen$ approach is based on a clear and rigorous physical picture, and allows a selective inclusion of specific higher-order contributions to the EPI, which is crucial to speed up the convergence. Moreover, our approach is suited to be used in conjunction with Hamiltonian learning methods~\cite{PhysRevLett.112.190501}. Indeed, the bottleneck of the method is the computation from first principles of the Kohn-Sham potential $\Vkshat(\bu)$ for several distorted atomic configurations $\bR=\bReq+\bu$, and the usage of machine learning techniques to infer these quantities using less first-principles calculations is definitely an approach that deserves to be explored. Another method to compute EPI beyond the lowest order that has recently aroused particular interest is the cumulant method, a way to include high-order terms in the electron self-energy, basically by exponentiating the AHC self-energy~\cite{PhysRevB.90.195135,PhysRevB.105.245120}.  Again, compared to the cumulant approach, our $\GWen$ method stands out for the conceptual clarity and rigour (e.g. about the \textit{ratio} behind the choice of the high-order perturbative diagrams considered), and for the full control of the approximation considered (i.e. a full control of what high-order terms are included). On the other side, the cumulant approach is not adiabatic, whereas $\GWen$ assumes instantaneous electronic screening (i.e. much faster than the nuclei screening). However, our novel approach can include non-adiabatic corrections together with non-linear and anharmonic effects, for instance in the calculation of the phonon-induced electronic band gap renormalization in semiconductors~\cite{miglio_predominance_2020}, by keeping the frequency dependence of the $\GWen$ self-energy, and thus essentially including non-linear effects in the so called non-adiabatic AHC theory~\cite{PhysRevB.18.5217,doi:10.1063/1.4927081,grimvall1981electron}. Therefore, we believe that our novel approach will not only have a large impact on strongly anharmonic superconductors, like the hydrogen-based ones, but also in the calculation of other properties related to the electron-phonon interaction such as electronic transport or the impact of lattice vibrations in the electronic bands. 

\section*{Acknowledgements}

We acknowledge fruitful discussions with Philip Allen,
Lorenzo Paulatto, Matteo Calandra, and Trinidad Novoa. We received financial support from the European Research Council (ERC) under the European
Union’s Horizon 2020 research and innovation program
(Grant Agreement No. 802533); 
the PID2022-142861NA-I00  project funded by MICIU/AEI/10.13039/501100011033 and FEDER, UE;
the Department
of Education, Universities and Research of the Eusko
Jaurlaritza and the University of the Basque Country UPV/EHU (Grant No. IT1527-22); and Simons Foundation through the Collaboration on New Frontiers in Superconductivity (Grant No. SFI-MPS-NFS-00006741-10). Computational resources were granted by the Red Española de Supercomputación (Grant No. FI-2022-3-0010).

%%%%%%%%%%%%%%%%%%%%%%%%%%%%%%%%%%%%%%%%%%%%%%%%%%%%%%%%%%%%%%%%%%%%%%%%%%%%%%%%%%%%%%%%%%%%%%%%%%%%

\appendix

%%%%%%%%%%%%%%%%%%%%%%%%%%%%%%%%%%%%%%%%%%%%%%%%%%%%%%%%%%%%%%%%%%%%%%%%%%%%%%%%%%%%%%%%%%%%%%%%%%%%%%%%%%%%5
\section{From $\Wen_{n'\bk'\,n\bk}(iw_l)$ to $\lambda_{n\bk}$}
\label{sec:from_W_to_lambda}

In this appendix we give more details about the passages that in Sec.~\ref{sec:GWen_approx} lead from $\Wen_{n'\bk'\,n\bk}(iw_l)$ to the formula Eq.~\eqref{eq:lambdank_T0} to compute $\lambda_{n\bk}$. The starting point is the formula for the self-energy, Eq.~\eqref{eq:slef_eph_bloch}, with the free-electron Green function $G_{n'\bk'}(ip_h-iw_l)$, Eq.~\eqref{eq:free_electron_G}. Since we consider a dirty/isotropic limit, we can neglect the details of $\Velph_{n'\bk'\,n\bk}(iw_l)$ w.r.t $n'\bk'$ on constant-energy surfaces and perform averages. The sum over $n'\bk'$ in 
Eq.~\eqref{eq:slef_eph_bloch} thus can be written as
\begin{align}
&\sum_{n'\bk'}\frac{1}{ip_h-iw_l-\varepsilon_{n'\bk'}}\,\Velph_{n'\bk'\,n\bk}(iw_l)\nonumber\\
% &\quad=\int d\varepsilon \frac{1}{ip_h-iw_l-\varepsilon}N(\varepsilon)\Velph_{\avg{n'\bk'}_{\varepsilon}\,n\bk}(iw_l)
&\quad=\Nc\int \frac{d\varepsilon }{ip_h-iw_l-\varepsilon}N(\varepsilon)
\Avg{\Velph_{n'\bk'\,n\bk}(iw_l)}_{\enex{n'\bk'}=\ene}
\label{eq:sum_n'k'}
\end{align}
where $N(\varepsilon)=\frac{1}{\Nc}\sum_{n\bk}\delta(\varepsilon-\varepsilon_{n\bk})$ 
is the density of states per spin (we are considering spin-degenerate case) and per unit cell, and 
\begin{equation}
\Avg{\Velph_{n'\bk'\,n\bk}(iw_l)}_{\enex{n'\bk'}=\varepsilon}
=\frac{\sum_{n'\bk'}\Velph_{n'\bk'\,n\bk}(iw_l)\delta(\varepsilon-\varepsilon_{n'\bk'})}
{\sum_{n'\bk'}\delta(\varepsilon-\varepsilon_{n'\bk'})}
\end{equation}
is the average of $\Velph_{n'\bk'\,n\bk}(iw_l)$ w.r.t $n'\bk'$ over the energy-constant surface
$\{n'\bk':\,\varepsilon_{n'\bk'}=\varepsilon \}$~(see App.~\ref{sec:sum_to_int} for more details).
In a low-temperature regime, most of the physics happen close to the Fermi level (typically within an energy shell around the Fermi level with a size of the order of the Debye frequency $\omegaD$). Therefore, in Eq.~\eqref{eq:slef_eph_bloch} we consider states $\Ket{n\bk}$ with energy $\varepsilon_{n\bk}$ within $~\pm\omegaD$, and values $p_{h}$ of the same order of magnitude. Since $w_{l}$ has this order too (Cfr.~Eqs.~\eqref{eq:Velph_Gauss_freq_normal_zeroT} and~\eqref{eq:Bsum_2} for the $T=0$ case), $G_{n'\bk'}(ip_h-iw_l)$ in Eq.~\eqref{eq:sum_n'k'} enters with small values of $|p_h-w_l|$, and therefore it is small for $|\varepsilon_{n'\bk'}|>\omegaD$. In conclusion, in Eq.~\eqref{eq:sum_n'k'} the major contributions to the result come from the sum over states $\Ket{n'\bk'}$ with energy $\varepsilon_{n'\bk'}$ within $(-\omegaD,+\omegaD)$, i.e.
from the integral in $d\varepsilon$ over the region $(-\omegaD,+\omegaD)$. 
The density of states $N(\varepsilon)$ and constant-energy surface average $\Avg{\Velph_{n'\bk'\,n\bk}(iw_l)}_{\enex{n'\bk'}=\varepsilon}$ are not expected to vary rapidly w.r.t. $\varepsilon$ within this narrow energy window, thus we can write
\begin{align}
&\sum_{n'\bk'}\frac{1}{ip_h-iw_l-\varepsilon_{n'\bk'}}\,\Velph_{n'\bk'\,n\bk}(iw_l)\nonumber\\
% &\quad=\int d\varepsilon \frac{1}{ip_h-iw_l-\varepsilon}N(\varepsilon)\Velph_{\avg{n'\bk'}_{\varepsilon}\,n\bk}(iw_l)
&\quad=\Nc\Nf\Velph_{n\bk}(iw_l)\int \frac{d\varepsilon }{ip_h-iw_l-\varepsilon}
\label{eq:sum_n'k'_ef}
\end{align}
where $\Nf=N(\ef)=N(0)$ is the Fermi-level density of states, and $\Velph_{n\bk}(iw_l)=
\avg{\Velph_{n'\bk'\,n\bk}(iw_l)}_{\enex{n'\bk'}=\ef}$ is the average of $\Velph_{n'\bk'\,n\bk}(iw_l)$ w.r.t $n'\bk'$ on the Fermi surface (FS), which we will indicate also with the notation $\avg{\Velph_{n'\bk'\,n\bk}(iw_l)}_{n'\bk'\in\FS}$.

From Eq.~\eqref{eq:Velph_w_a2F}, performing the Fermi-surface average w.r.t $n'\bk'$ we have
\begin{align}
\Velph_{n\bk}(iw_l)&=\frac{1}{\Nc\Nf}\int_{0}^{+\infty}d\omega\,\asqF_{n\bk}(\omega)\nonumber\\
&\qquad\qquad  \left[\frac{1}{iw_l-\omega}-\frac{1}{iw_l+\omega}\right]\,,
\label{eq:spectralWenavg}
\end{align}
where $\asqF_{n\bk}(\omega)=\Avg{\asqF_{n'\bk'\,n\bk}(\omega)}_{n'\bk'\in \FS}$. Plugging this formula in Eq.~\eqref{eq:sum_n'k'_ef} , from Eq.~\eqref{eq:slef_eph_bloch} we obtain the formula Eq.~\eqref{eq:sigmaep_G}
for the self-energy, with 
% \begin{equation}
% \Sigma_{n\bk}(ip_h)=\int_0^{+\infty}d\omega\asqF_{n\bk}(\omega)\int_{-\infty}^{+\infty}d\varepsilon I[ip_h,\varepsilon,\omega]\,,
% \label{eq:sigmaep_G}
% \end{equation}
\begin{align}
I[ip_h,\varepsilon,\omega]&=-\frac{1}{\beta}\sum_l\left[\frac{1}{ip_h-iw_l-\varepsilon}\right]\nonumber\\
&\mkern100mu\left[\frac{1}{iw_l-\omega}-\frac{1}{iw_l+\omega}\right]\\
&=\left[\frac{1-f(\varepsilon)+n(\omega)}{ip_h-\varepsilon-\omega}+\frac{f(\varepsilon)+n(\omega)}{ip_h-\varepsilon+\omega}\right]
\label{eq:Ifunc}\\
&=\left[\frac{\Theta(\varepsilon)}{ip_h-\varepsilon-\omega}+\frac{1-\Theta(\varepsilon)}{ip_h-\varepsilon+\omega}\right]\,,
\label{eq:IfuncT0}
\end{align}
where in the second line we have done the Matsubara frequency summation, with $f(x)=1/(e^{\beta x}+1)$ and $n(x)=1/(e^{\beta x}-1)$ the Fermi-Dirac and the Bose-Einstein occupation functions, respectively, and in the third line, compatibly with the fact that we are considering a low-temperature case, we have considered the zero-temperature expression of
these occupation functions, $n(\omega)=0$ and $f(\varepsilon)=1-\Theta(\varepsilon)$, where $\Theta(\varepsilon)$ is the Heaviside's step function. Eqs.~\eqref{eq:sigmaep_G} and~\eqref{eq:I} are thus demonstrated. From Eq.~\eqref{eq:I}, it is
\begin{align}
&
\left.-\frac{\partial}{\partial \Omega}\Re\left(\int_{-\infty}^{+\infty}\,d\varepsilon\,I[\Omega+i0^+,\varepsilon,\omega]\right)\right|_{\Omega=0}
\nonumber\\
&\qquad=\int_{0}^{+\infty}\,\frac{d\varepsilon}{(\varepsilon+\omega)^2}+\int_{-\infty}^{0}\,\frac{d\varepsilon}{(\varepsilon-\omega)^2}\\
&\qquad=\frac{2}{\omega}\,
\end{align}
thus Eq.~\eqref{eq:lambdank_T0} is readily demonstrated. 
%%%%%%%%%%%%%%%%%%%%%%%%%%%%%%%%%%%%%%%%%%%%%%%%%%%%%%%%%%%%%%%%%%%%%

%%%%%%%%%%%%%%%%%%%%%%%%%%%%%%%%%%%%%%%%%%%%%%%%%%%%%%%%%%%%%%%%%%%%
\section{Derivation Eq.~\eqref{eq:sum_n'k'}}
\label{sec:sum_to_int}
In this appendix we give more details about the derivation of Eq.~\eqref{eq:sum_n'k'}. 
In the thermodynamic limit ($N_c\gg 1$) is
\begin{align}
\frac{1}{\Nc}\sum_{n\bk}
&=\sum_{n}\frac{1}{\Omega_{BZ}}\int_{BZ}d\bk\\
&=\sum_{n}\frac{1}{\Omega_{BZ}}
\int d\varepsilon \int_{\enex{n\bk}={\varepsilon}}\frac{dS_{\bk}}{|\nabla_{\bk}\varepsilon_{n\bk}|}\,,
\end{align}
where $N_c$ is the number of unit cells comprising the supercell with periodic boundary conditions, equal to 
the number of considered $\bk$ points fulfilling the corresponding Born-Von Karman conditions, and the $dS_{\bk}$ integral is a surface integral in the $\bk$ space, performed over the constant-energy surface  $\{\bk:\,\enex{n\bk}={\varepsilon}\}$. Notice that the density of states per unit cell given by the $n$-th band $N_n(\varepsilon)$ is
\begin{align}
N_n(\varepsilon)&=\frac{1}{\Nc}\sum_{\bk}\delta(\varepsilon-\enex{n\bk})\\
&=\frac{1}{\Omega_{BZ}}\int d\bk\,\delta(\varepsilon-\enex{n\bk})\\
&=\frac{1}{\Omega_{BZ}}\int d\varepsilon' \int_{\enex{n\bk=\varepsilon'}} 
\frac{dS_{\bk}}{|\nabla_{\bk}\varepsilon_{n\bk}|}\,\delta(\varepsilon-\enex{n\bk})\\
&=\frac{1}{\Omega_{BZ}}\int d\varepsilon' \,\delta(\varepsilon-\varepsilon')\int_{\enex{n\bk=\varepsilon'}} 
\frac{dS_{\bk}}{|\nabla_{\bk}\varepsilon_{n\bk}|}\\
&=\frac{1}{\Omega_{BZ}}\int_{\enex{n\bk=\varepsilon}} 
\frac{dS_{\bk}}{|\nabla_{\bk}\varepsilon_{n\bk}|}\,.
\end{align}
For each $n\bk$, we consider $\Avg{\Velph_{n'\bk'\,n\bk}(iw_l)}_{\varepsilon_{n'\bk'}=\varepsilon}$,
the average of $\Velph_{n'\bk'\,n\bk}(iw_l)$ with respect to $n'\bk'$ over 
the constant-energy surface $\{n'\bk':\,\enex{n'\bk'}=\varepsilon\}$:
\begin{align}
&\Avg{\Velph_{n'\bk'\,n\bk}(iw_l)}_{\varepsilon_{n'\bk'}=\varepsilon}\nonumber\\
&\qquad=\frac{\sum_{n'\bk'} \Velph_{n'\bk'\,n\bk}(iw_l) \delta(\enex{n'\bk'}-\varepsilon) }{\sum_{n'\bk'}\delta(\enex{n'\bk'}-\varepsilon)}\\
&\qquad=\frac{\sum_{n'}\int_{\enex{n'\bk'}=\varepsilon} 
\frac{dS_{\bk'}}{|\nabla_{\bk'}\varepsilon_{n'\bk'}|}\Velph_{n'\bk'\,n\bk}(iw_l)}
{\sum_{n'}\int_{\enex{n'\bk'}=\varepsilon} 
\frac{dS_{\bk'}}{|\nabla_{\bk'}\varepsilon_{n'\bk'}|}}\,.
\end{align}
Notice that, by definition, $\Avg{\Velph_{n'\bk'\,n\bk}(iw_l)}_{\varepsilon_{n'\bk'}=\varepsilon}$ 
does not depend on $n',\bk'$ and it is
\begin{align}
&\sum_{n'\bk'}\delta(\ene-\enex{n'\bk'})\Avg{\Velph_{n'\bk'\,n\bk}(iw_l)}_{\varepsilon_{n'\bk'}=\varepsilon}
\nonumber\\
&
\mkern80mu=
\sum_{n'\bk'}\delta(\ene-\enex{n'\bk'})\Velph_{n'\bk'\,n\bk}(iw_l)\,,
\end{align}
or, in the thermodynamic limit,
\begin{align}
&\sum_{n'}\int_{\enex{n'\bk'}=\varepsilon} 
\frac{dS_{\bk'}}{|\nabla_{\bk'}\varepsilon_{n'\bk'}|}
\Avg{\Velph_{n'\bk'\,n\bk}(iw_l)}_{\varepsilon_{n'\bk'}=\varepsilon}
=\nonumber\\
&\mkern80mu=\sum_{n'}\int_{\enex{n'\bk'}=\varepsilon} 
\frac{dS_{\bk'}}{|\nabla_{\bk'}\varepsilon_{n'\bk'}|}\Velph_{n'\bk'\,n\bk}(iw_l)\,.
\end{align}
Assuming that, given $n\bk$, we can neglect the dependence of $\Velph_{n'\bk'\,n\bk}(iw_l)$
w.r.t. $n'\bk'$ on the constant-energy surface $\{n'\bk':\,\enex{n'\bk'}=\varepsilon\}$ , 
and thus replace $\Velph_{n'\bk'\,n\bk}(iw_l)$ with its surface average
$\Avg{\Velph_{n'\bk' n\bk}(iw_l)}_{\varepsilon_{n'\bk'}=\varepsilon}$, we have
\begin{align}
&\sum_{n'\bk'}\frac{1}{ip_h-iw_l-\varepsilon_{n'\bk'}}\,\Velph_{n'\bk' n\bk}(iw_l)=\\
&=\sum_{n'}\frac{N_c}{\Omega_{BZ}}
\int d\varepsilon
\frac{1}{ip_h-iw_l-\varepsilon}\nonumber\\
&\mkern100mu\times\int_{\varepsilon_{n'\bk'}=\varepsilon}
\frac{dS_{\bk'}}{|\nabla_{\bk'}\varepsilon_{n'\bk'}|}
\Velph_{n'\bk'\,n\bk}(iw_l)\\
&=\Nc\int d\varepsilon
\frac{1}{ip_h-iw_l-\varepsilon}\nonumber\\
&\mkern100mu\times\sum_{n'}\frac{1}{\Omega_{BZ}}\int_{\varepsilon_{n'\bk'}=\varepsilon}
\frac{dS_{\bk'}}{|\nabla_{\bk'}\varepsilon_{n'\bk'}|}
\Velph_{n'\bk'\,n\bk}(iw_l)\\
&=\Nc\int d\varepsilon
\frac{1}{ip_h-iw_l-\varepsilon}\Avg{\Velph_{n'\bk'\,n\bk}(iw_l)}_{\varepsilon_{n'\bk'}=\varepsilon}\nonumber\\
&\mkern100mu\times\sum_{n'}\frac{1}{\Omega_{BZ}}
\int_{\varepsilon_{n'\bk'}=\varepsilon}\frac{dS_{\bk'}}{|\nabla_{\bk'}\varepsilon_{n'\bk'}|}\\
&=\Nc\int d\varepsilon
\frac{1}{ip_h-iw_l-\varepsilon}\Avg{\Velph_{n'\bk'\,n\bk}(iw_l)}_{\varepsilon_{n'\bk'}=\varepsilon}\nonumber\\
&\mkern100mu\times\sum_{n'}N_{n'}(\varepsilon)\\
&=\Nc\int d\varepsilon
\frac{N(\varepsilon)}{ip_h-iw_l-\varepsilon}\Avg{\Velph_{n'\bk'\,n\bk}(iw_l)}_{\varepsilon_{n'\bk'}=\varepsilon}\,,
\end{align}
which demonstrates Eq.~\eqref{eq:sum_n'k'}.
\section{Restriced sum to avoid numerical instabilities}
\label{sec:restricted_sum}
In this appendix we want to explain with more details why neglecting the $\bk=0$ and $n=m$ terms in Eq.~\eqref{eq:eli_impl} is convenient to avoid numerical instabilities. 
The problem essentially stems from the presence of a double Dirac delta in Eq.~\eqref{eq:eli_impl}, which is a mathematically ill-defined object.
In standard unit-cell calculations, this issue is routinely addressed in the evaluation of the electron–phonon coupling—which involves sums of the form $\sum_{\bq \bk}\sum_{nm}\delta\!\left(\enex{\bk+\bq n}-\ef\right)\delta\!\left(\enex{\bk m}-\ef\right)$—by explicitly excluding the $\bq=0$ contribution (i.e., the $\Gamma$ point). From this expression, it is also evident that the singularity arising from the $\bq=0$ and $n=m$ terms is integrable, i.e., it yields a finite result. Indeed, by performing a simple change of variables, it can be shown that the double sum corresponds to the square of the density of states at the Fermi level:
\begin{equation}
\sum_{\bq  \bk}\sum_{n m}\delta\!\left(\enex{\bk+\bq n}-\ef\right)\delta\!\left(\enex{\bk m}-\ef\right)
=\left(\sum_{\bk n}\delta\!\left(\enex{\bk n}-\ef\right)\right)^2
\end{equation}
The exclusion of the $\bq=0$ point is therefore employed only to speed up the convergence with respect to the $\bq$ integration.
However, this strategy cannot be applied in real-space supercell calculations, where only $\Gamma$-point distortions are considered, and the convergence with respect to the $\bq$ integration corresponds instead to the convergence with respect to the supercell size. In this case, in order to avoid numerical instabilities that could hamper the convergence, in  Eq.~\eqref{eq:eli_impl} the diagonal terms $n=m$ terms are excluded from the sum. Of course, excluding only the ``diagonal terms'' $n=m$ from the sum would affect the gauge invariance of the result in case of degenerate states around the Fermi level. For this reason, a proper way to introduce the restricted sum is to compute
\begin{equation}
\sum_{\bk n m}{}^{\!\!'}\equiv \sum_{\substack{\bk  n  m\\ \,\enex{\bk  n}\neq\enex{\bk  m}}}\,.
\end{equation}
Another useful way to look at the convergence issues of Eq.~\eqref{eq:eli_impl} is to rewrite the sum in the square brackets as follows:
\begin{align}
&\frac{1}{\Nc}\sum_{\bk\,n\,m}\delta\left(\epsilon_{n\,\bk}-\ef\right)
\delta\left(\epsilon_{m\bk}-\ef\right)   
\left|\gavgn^{mn\bk}_{\mu_1\ldots \mu_n}\right|^2\\
&=\frac{1}{\Omega_{BZ}}\sum_{n,m}\int_{BZ} d\bk\, \delta\left(\epsilon_{m\bk}-\ef\right)
\delta\left(\epsilon_{n\bk}-\ef\right)\left|\gavgn^{mn\bk}_{\mu_1\ldots \mu_n}\right|^2\\
&=\frac{1}{\Omega_{BZ}}\sum_{n,m}\int d\varepsilon \int d\varepsilon' 
\int_{\{\epsilon_{n\bk}=\varepsilon\}\cap \{\epsilon_{m\bk}=\varepsilon'\}} 
\nonumber\\
&\mkern10mu \times
\frac{d\ell_{\bk}}{|\nabla_{\bk}\enex{m\bk}\times \nabla_{\bk}\enex{n\bk}|}
\delta\left(\epsilon_{m\bk}-\ef\right)
\delta\left(\epsilon_{n\bk}-\ef\right)\left|\gavgn^{mn\bk}_{\mu_1\ldots \mu_n}\right|^2\\
&=\frac{1}{\Omega_{BZ}}\sum_{n,m}\int_{\{\epsilon_{n\bk}=\ef\}\cap \{\epsilon_{n\bk}=\ef\}} 
d\ell_{\bk}\,\frac{\left|\gavgn^{mn\bk}_{\mu_1\ldots \mu_n}\right|^2}{|v_{m\bk}\times v_{n\bk}|}\,,
\end{align}
where the $d\ell_{\bk}$ integral is a line integral in $\bk$, and $v_{n\bk}=\nabla_{\bk}\enex{n\bk}$ is the electron velocity. The integrand is divergent, because the denominator is zero, not only when $m=n$ but also when $\bk=0$ (if the system is time-reversal symmetric, in $\bk=0$ it is $\nabla_{\bk}\enex{n\bk}=0$). This explains why the $\bk=0$ term needs to be excluded too.
Therefore we write:
\begin{equation}
\sum_{\bk n m}{}^{\!\!'}=\sum_{\bk\neq 0}\sum_{\substack{n  m\\ \,\enex{\bk n}\neq\enex{\bk m}}}\,.
\end{equation}

%\bibliography{biblio}

\end{document}